\documentclass[reprint,amsmath,amssymb,aps,groupedaddress]{revtex4-1}

\usepackage[version=3]{mhchem} 
\usepackage[T1]{fontenc}       
\usepackage{amsmath, amssymb}
\usepackage{graphicx}
\usepackage{multirow}
\usepackage{array}
\usepackage{booktabs}
\usepackage{colortbl}
\usepackage{color}
\usepackage{soul}
\usepackage[normalem]{ulem}

\begin{document}

\author{Y.~Korneeva}
\affiliation{Physics Department, Moscow State University of Education, Moscow, Russia}
\email{korneeva@rplab.ru}
\author{D.Yu.~Vodolazov}
\affiliation{Institute for Physics of Microstructures, Russian Academy of Sciences, Nizhny Novgorod, GSP-105, Russia \\ and
Physics Department, Moscow State University of Education, Moscow, Russia}
\author{A.~V.~Semenov}
\affiliation{Physics Department, Moscow State University of Education, Moscow, Russia\\ and
Moscow Institute of Physics and Technology (State University), Moscow, Russia}
\author{I.~Florya}
\affiliation{Physics Department, Moscow State University of Education, Moscow, Russia }
\author{E.~Baeva}
\affiliation{Higher School of Economics National Research University,  Moscow, Russia \\ and
Physics Department, Moscow State University of Education, Moscow, Russia}
\author{N.~Simonov}
\affiliation{Physics Department, Moscow State University of Education, Moscow, Russia }
\author{A.A.~Korneev}
\affiliation{Physics Department, Moscow State University of Education, Moscow, Russia,\\
Moscow Institute of Physics and Technology (State University), Moscow, Russia\\ and
Higher School of Economics National Research University,  Moscow, Russia}
\author{G.N.~Goltsman}
\affiliation{Physics Department, Moscow State University of Education, Moscow, Russia\\ and
Higher School of Economics National Research University,  Moscow, Russia}
\author{T.M.~Klapwijk}
\affiliation{Physics Department, Moscow State University of Education, Moscow, Russia\\ and
Kavli Institute of Nanoscience, Delft University of Technology, The Netherlands}

\date{\today}

\title  {Optical single photon detection in micron-scaled NbN bridges}


\begin{abstract}
We demonstrate experimentally that single photon detection can be achieved in micron-wide NbN bridges, with widths ranging from  0.53\,$\mu$m to 5.15\,$\mu$m and for photon-wavelengths
from 408\,nm to 1550\,nm. The microbridges are biased  with a dc current close to the experimental critical current, which is estimated to be about 50\% of
the theoretically expected depairing current. These results offer an alternative to the standard superconducting single-photon detectors (SSPDs), based on nanometer scale nanowires implemented in a long meandering structure.  The results are consistent with improved theoretical modelling based on the theory of non-equilibrium superconductivity including the vortex-assisted mechanism of initial dissipation.
\end{abstract}

\maketitle

\section{Introduction}
The present superconducting nanowire single photon detectors
(SSPD's) are based on long meandering superconducting strips with
a width in the range of 50 to 150\,nm~\cite{NatarajanSUST2012}. It
has been empirically found that the use of wider strips leads
either to the loss of the single photon nature of the response or
to a rather small detection efficiency
\cite{MaingaultJAP10,LuscheJAP2014}. This result is in line with
the initial interpretation of this type of detector
\cite{SemenovPhysicaC2001,SemenovEurPhysJ2005}, in which it was
understood that the width of the supercurrent-carrying strip
should be comparable to the diameter, $D$, of the normal
hot spot (a region where the superconducting state is suppressed),
due to the absorption of the photon. If the strip is biased near
its experimentally determined critical current, the emergence of
the hot-spot forces a redistribution of the supercurrent, leading
to a locally enhanced supercurrent-density, triggering the switch
to the resistive state.  Using simple estimates, based on the
conservation of energy and typical parameters for niobium-nitride
(NbN), given the  energy of an optical photon leads to $D
\sim40$\,nm for the expected size of the normal hot spot
\cite{Ourbook}.

This geometrical mechanism to exceed the critical current-density,
has initiated a more thorough analysis of the conditions of the
superconducting strips under current-bias by Zotova and Vodolazov
\cite{ZotovaPRB2012, VodolazovPRA2017}.  They consider a
superconducting strip, biased sufficiently close to the intrinsic {\it
depairing} current $I_{dep}$.  Then a small amount of energy can switch the current-carrying superconductor to a resistive state, 
with the needed energy going to zero when $I$\ approaches $I_{dep}$.The only requirement
for the width of the strip is that it should be smaller than the
Pearl penetration depth $\Lambda=2\lambda^2/d$ (with $\lambda$ the
London penetration depth for a dirty superconductor and $d$ the
thickness of the strip). Under these conditions the supercurrent will be uniform across the width, whereas for wider strips the supercurrent will 
be distributed non-uniformly.  For a typical NbN films with a thickness about
5\,nm and $\lambda \sim 470$\,nm \cite{lambdaAPL2010} one obtains
$\Lambda \simeq 90$\, $\mu$m.

Using the microscopic theory for superconductivity it was
shown in Refs.~\citenum{ZotovaPRB2012, VodolazovPRA2017}, that if such a
strip with a uniform supercurrent $I$ can be biased at $I \gtrsim
0.5-0.7 I_{dep}$ the superconducting state becomes unstable in
response to relatively small additions of energy in the form of a localized disturbance, loosely called a 'hot spot'. Its specific nature in terms of the microscopic theory of non-equilibrium superconductivity has not yet been worked out. It is considered to be a localized nonequilibrium distribution over the energies and with at least a depressed local energy gap initially surrounded by an equilibrium superconductor. The dynamics of such an 'impact crater' in the superconducting film will depend on the materials. 

In previous work this process was mostly described by what we label as a 'geometric-hot-spot-model'. The essential feature of this approach is that the supercurrent, initially carried over the full width $w$ of the supercurrent in the wire is pushed to a more narrow part, excluding the 'hot' part with a diameter $d$. This increased current density then may exceed the critical current density initiating a transition to a voltage-carrying state. To optimize the efficiency of detection the wire should be of the order of the size $d$ of the hot spot in the superconductor created by the absorbed photon. This geometric hot spot model is often used for a qualitative discussions and has been mostly leading the technological development. 

More recently, the microscopic approach has emerged including the use of nonequilibrium superconductivity. In this approach the phase-coherence of the superfluid flow is fully taken into account, as well as the emergence of resistivity in the superconductor by the creation of vortices. In order to make a clear distinction with previous approaches we call this the 'photon-generated superconducting vortex model'. The theory states that the efficiency of the photon-detection is not determined by the geometry, as long as the initial current-density is uniform and close to the critical pair-breaking current. The requirement for uniformity of the current density is given in a previous paragraph. If the superconducting wire can be biased close to the critical pair-breaking current, all photons will be detected, wherever they hit the wire, because all of them create a sufficient disturbance to trigger a \emph{local} excess of the critical current density, initiating the creation of vortex-anti-vortex pairs.
In the film vortex-antivortex pairs will be created inside the
hot spot (if it is located far from the edge of the
strip) or by vortex entry into the strip (if the hot spot 
is located close to the edge). The motion of a vortex and/or
anti-vortex due to the Lorentz force leads to a voltage in the
superconductor and eventually to the appearance of a normal domain
\cite{VodolazovPRA2017}. Rather than assuming a fully normal hot
spot this model takes into account the resistive properties of the
superconducting state due to vortex-movement, with details
determining the full dynamics.

In order to build experimentally on a model based
on current densities close to the critical pair-breaking current
one needs to determine whether the observed critical current is
determined by intrinsic or extrinsic properties, such as material
imperfections. The commonly used material for single
photon-detection is NbN with a thickness of about 5 nanometer,
chosen because of its fast electron-phonon relaxation including
phonon-escape to the substrate. However, such films have a fairly
high resistivity, a low diffusion constant, and a high resistance
per square in the order of $800~\Omega$, which implies that they
have an intrinsic tendency to become electronically inhomogeneous
with a spatially fluctuating superconducting energy gap \cite{KamlapureSciRep2013,HortensiusIEEE2013}, which may worsen due to material
imperfections. Therefore, it is to be expected that the critical
current for a long superconducting wire is determined by the
weakest spot, which statistically will be the lowest value of the
energy gap along the wire. Secondly, in order to know the value of
the critical depairing current one has to rely on a quantitative
estimate based on measured parameters, and preferably on a
comparison with the functional dependence, such as carried out for
aluminium by Romijn et al \cite{RomijnPRB1982}. Relatively high
critical current densities have been reported recently by Charaev
et al \cite{Charaev2017}, although in more narrow strips larger
values have been reported \cite{LuscheJAP2014,IliinPRB14}.
Nevertheless, the maximally reachable value is not known and,
given the expected inhomogeneities, may vary. Instead we have
decided to work with relatively short microbridges and vary the
width, to minimize the risks of hitting a too low critical current
while expecting a reasonably uniform current density.

In this work we report on our findings that relatively wide NbN bridges with
widths in the range of 0.53\,$\mu$m to 5.15\,$\mu$m are able to
detect single photons of wavelengths from 408\,nm to 1550\,nm. We
determine the Internal Detection Efficiency, $IDE$, the detection
efficiency normalized to the absorption, for different bias
currents $I$,  and find that it reaches value of tens of percents near the experimental critical
current $I_c$.
From the experimental data we distinguish two regimes.
\textit{Regime I} in which for increasing current a sharp increase
in $IDE$ is observed, analogous to the
conventional meanderlike SSPDs. It is followed by a much slower
increase of the $IDE$  upon approaching $I_c$, which we label \textit{Regime II}. We attribute
\textit{Regime I} to fluctuation assisted photon detection with
the slope of the $IDE$ as a function of current $I$ comparable to
the slope of the number of dark counts with the bias current. The
\textit{Regime II} we attribute to deterministic photon counting as described theoretically by Zotova and Vodolazov \cite{ZotovaSUST2014}. We are able to explain quantitatively the dependence of the $IDE$ on the current $I$ in \textit{Regime II} for the short wavelengths, taking into account the actual geometry of our samples. We argue that in this parameter-range we observe an $IDE$ close to unity.

Our findings offer a new route towards superconducting single photon detectors
with a short dead time, a few hundreds of
picoseconds, because of the relatively small kinetic inductance
of short superconducting bridges in
comparison to the conventional superconducting meanders \cite{NatarajanSUST2012}. 
Additionally,  our
results provide support for the relevance of the vortex-assisted contribution to
photon detection as proposed by one of the authors  \cite{VodolazovPRA2017}.

\section{Samples and characterization}

\begin{table*}[t]
  \caption{Parameters of the studied samples for temperature $T$=4.2\,K. Width of the bridge $w$ is at the neck, $T_c$ is the critical temperature determined from the midpoint of the resistive transition,
$\rho(20K)$ is resistivity at $T$=20\,K, $j_c$ and $j_c^{sh}$ are the critical current densities measured without and with a shunt resistor, $j_{dep}$ is the calculated depairing current at the indicated temperature, using $j_{dep}(0)$ the calculated critical depairing current at $T=0$ following from Eq.\ref{Eq:Idep0} The variation in the calculated values are due to the variations in resistance per square determined for each sample. The diffusion constant $D$ is kept constant.}
  \label{tbl:Table1}
  \begin{ruledtabular}
  \begin{tabular}{llllllll}
    \hline
   Sample & width & $T_c$  & $\rho(20K)$ & $j_{c}(4.2 K)$  &  $j_{c}^{sh} (4.2 K)$ & ${j_{dep}}(4.2 K)$ & $j_{dep}(0)$ \\


 ID & $\mu$m & $K$ & $\mu\Omega\cdot cm$ & $A/cm^{2}$ & $A/cm^{2}$ & $A/cm^{2}$ & $A/cm^{2}$ \\
  \hline

$A$ & 0.53  &  8.25 &  386 & $3.16\cdot10^6$ & $3.67\cdot10^6$& $3.79\cdot10^6$ & $5.94\cdot10^6$ \\
$B$ & 1.61  &8.35& 396&  $2.74\cdot10^6$ & $3.72\cdot10^6$&$3.81\cdot10^6$ & $5.89\cdot10^6$ \\
$C$ & 2.12   &8.5& 393&  $3.75\cdot10^6$ & $4.43\cdot10^6$ &$4.02\cdot10^6$ & $6.11\cdot10^6$ \\
$D$ & 3.07   &8.35& 398&  $3.06\cdot10^6$ & $3.66\cdot10^6$ & $3.79\cdot10^6$ & $5.87\cdot10^6$ \\
$E$ & 4.04   &8.35& 402&  $2.52\cdot10^6$ &$3.16\cdot10^6$& $3.75\cdot10^6$ & $5.8\cdot10^6$ \\
$F$ & 5.15   &8.35& 427& $2.28\cdot10^6$ &$2.57\cdot10^6$& $3.54\cdot10^6$ & $5.47\cdot10^6$ \\
   \hline
  \end{tabular}
  \end{ruledtabular}
\end{table*}

\begin{figure}
    \includegraphics[width=0.5\textwidth]  {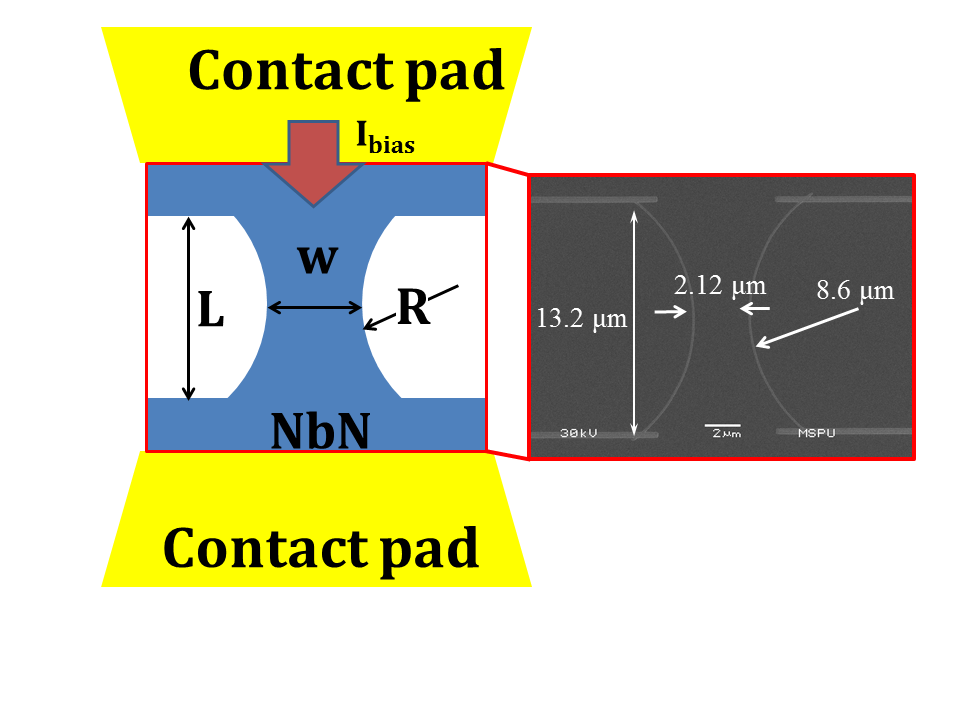}
    \caption {~Drawing of a typical NbN constriction-type bridge with a scanning electron microscope (SEM) image of one of the bridges with indicated dimensions (Sample $C$ in Table \ref{tbl:Table1}). The contacts on top of the NbN film are made of gold (Au). All bridges have edges designed as a segment of a circle with radius 8.6 $\mu m$.}
\label{fig:View}
\end{figure}

Our samples are planar constriction-type microbridges as shown in
Fig.~\ref{fig:View}. They are made from reactively sputtered NbN
films with a thickness, determined from a calibrated  sputter rate
and the sputter time,  of 5.8 nm. The width $w$ of the bridges at
the narrowest point was varied from 0.53\,$\mu$m to 5.15\,$\mu$m.
The constriction-type topology was chosen to prevent current
crowding effects~\cite{ClemBerggrenPRB2011} and to maximize the
chance to reach in the experiments the critical pair-breaking
current. Note that given the properties of the NbN films the constrictions are much larger in length and width than the coherence length. Therefore, it is assumed that the wider part, with
the lower current density does not lead to an enhancement of the
critical pair-breaking current in contrast to the case treated by Aslamazov and Larkin~\cite{Aslamazov69}   because of the short coherence length in NbN~\cite{Likharev79}. The parameters of the studied devices are summarized in Table \ref{tbl:Table1}. The details of the
fabrication process are presented in Appendix A.

\begin{figure}
    \centering
    \begin{tabular}{l}
        a)\\
    \includegraphics[width=8cm]{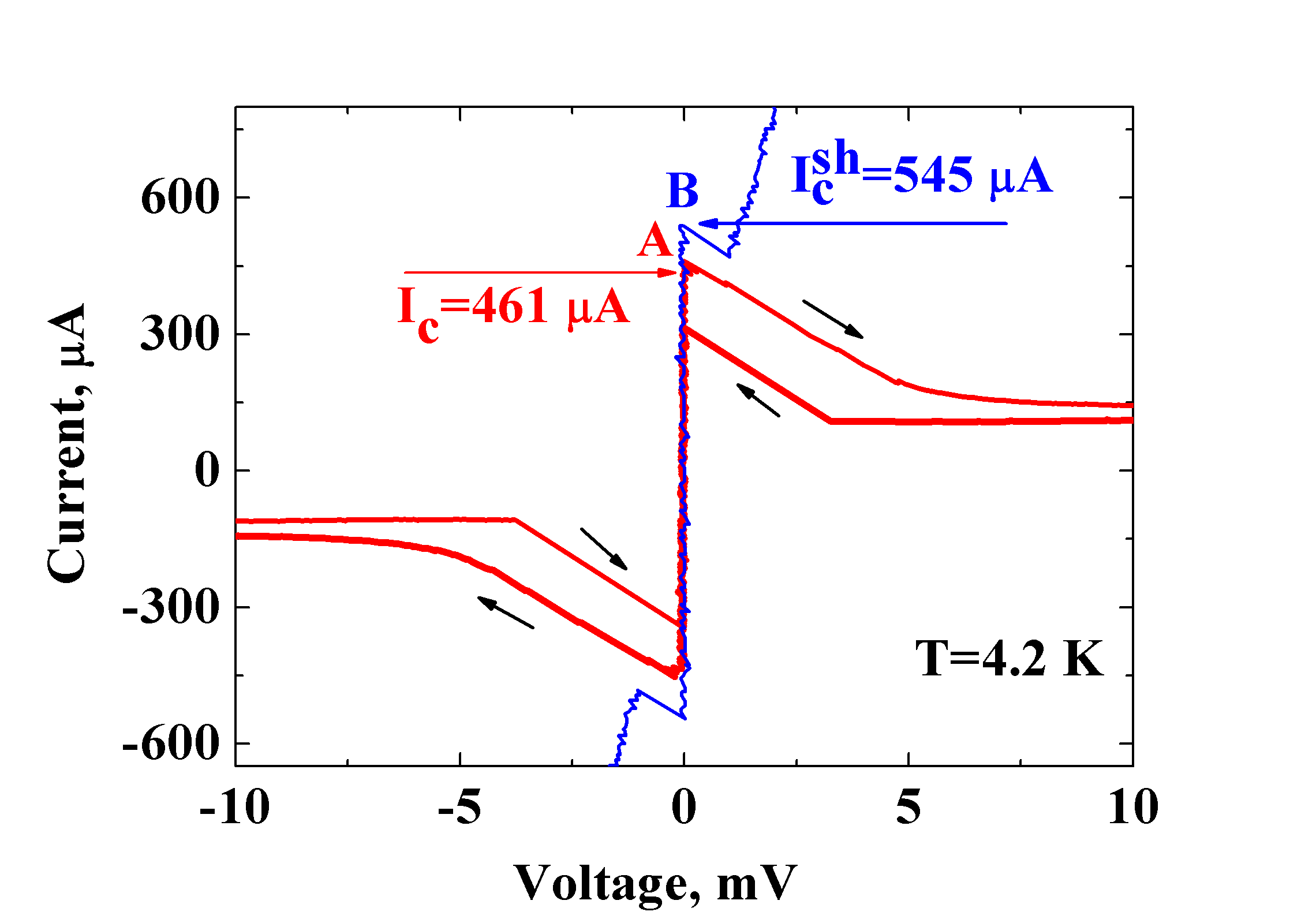} \\
            b)\\
        \includegraphics[width=8cm] {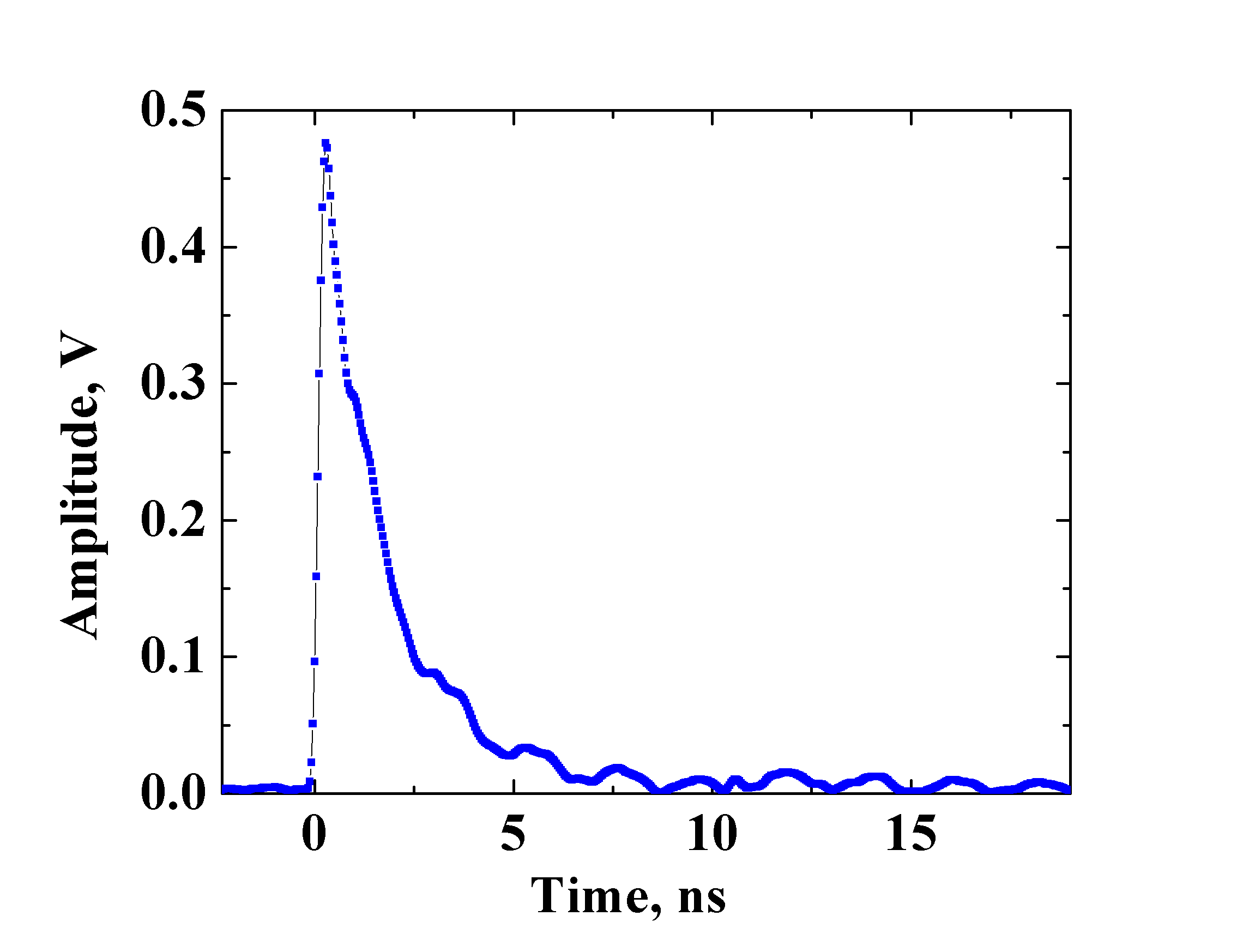}\\
    \end{tabular}
    \caption{(a)~I-V curves of Sample $C$ measured with 3\,$\Omega$ shunt resistor (blue) and without (red) at temperature $T$=4.2\,K. (b) Single-shot waveform transient from Sample C when a photon is absorbed.}
    \label{fig:Figure2}
\end{figure}

The experimentally observed critical current densities $j_c$ for two different temperatures
$T$=4.2\,K and $T$=1.7\,K.
In Table \ref{tbl:Table1} these results are listed for temperature $T$=4.2\,K. Obviously, there is
some scatter in the values of the critical temperature as well as
in the critical current density. It indicates that there is some
uncontrolled variation from sample to sample, although the samples
are from the same film. This may be caused by the metallurgy resulting from the
deposition-conditions but it may also be intrinsic due to the
competition between localisation in this low diffusivity material
and superconductivity. For a material like NbN the experimental values in this experimental geometry are reassuringly close to the theoretical values. Since these relatively wide samples have impractically high critical
currents we connect a 3 $\Omega$ shunt resistor in
parallel to the sample to prevent latching and to enable a
spontaneous return to the superconducting state. Although not expected, this leads to
different increases of experimentally observed critical currents when the shunt is connected. In Fig.~\ref{fig:Figure2}(a) we show current-voltage characteristics for Sample C without (red curve) and with the 3 $\Omega$ shunt resistor (blue curve) with different critical currents  (point A and
point B, respectively). A similar observation
has been reported and discussed previously by Brenner et al \cite{Brener2012}. 
In our measurements this can be attributed, at least in part, to division of the current between the chip with superconducting bridge and the shunt: the branch with superconducting bridge contains also normal resistance, of order of a few tenths of Ohms, because this is a two-point measurement. Of course, the quantity of interest is a supercurrent, \emph{i.e.} the current flowing through the sample; hence, we use the measured current with the shunt only as a value to plot the data.  

We assume that the theoretical depairing currents can be described
by the expression derived for clean superconductors by Bardeen
~\cite{BardeenRevModPhys1962}. It deviates by less than 3$\%$ from
the results of the microscopic calculations for dirty
superconductors, and which have been compared to experiments on
aluminium by Romijn et al \cite{RomijnPRB1982}):
\begin{equation}
I_{dep}(T)=I_{dep}(0)\left[ 1- \left( \frac{T}{T_c} \right)^2 \right]^{3/2}
\label{Eq:Idep}
\end{equation}
with the prefactor $I_{dep}(0)$, calculated from Eq.\,31 in Clem et al \cite{ClemPRB2012v86}:
\begin{equation}
I_{dep}(0)=0.74\frac{w\left[ \Delta(0)\right]^{3/2}}{e R_s \sqrt{\hbar D}},
\label{Eq:Idep0}
\end{equation}
Here, $\Delta(0)$ is the superconducting energy gap at 0\,K, $e$ is
the electron charge, $R_s$ is the resistance per square, $D$ is the diffusivity.
Strictly speaking Eqs.(\ref{Eq:Idep},\ref{Eq:Idep0}) are
quantitatively valid for moderately dirty superconductors with
$k_F l \gg 1$ ($k_F$ is the Fermi wavevector and $l$ is the mean
free path for elastic scattering). We also assume the BCS-ratio of $\Delta(0)/ k_B T_c \approx 1.76$.

In applying these expressions to the present NbN films we make a conceptual step, which would require a deeper justification, and which is currently not available. It is known that the critical temperature $T_c$ varies with the resistance per square, reminiscent of experiments on the superconductor-insulator transition\cite{lambdaAPL2010,WangJAP1996,ChokalingamPRB2009,BeckPRL2011,NoatPRB2013}. It has also been found that the ratio of $\Delta(0)/ k_BT_c$ for such films is not a constant but changes with the resistance per square. It is often attributed to the film properties and the substrate
surface~\cite{RomestainNewJPhys2004}. Nevertheless, there is compelling evidence that these materials are anomalous in many respects~\cite{arxivKapitulnik2017}. Given this uncertainty we make the choice to take the most straightforward input towards  Eqs.(\ref{Eq:Idep},\ref{Eq:Idep0}) and we use the BCS ratio for $\Delta$ and $k_BT_c$ and for $k_F l
\approx$3--5~\cite{RomestainNewJPhys2004, ChokalingamPRB2009,
SemenovPRB2009, NoatPRB2013}.
For the diffusion contant we use $D$=0.31\,cm$^2$/s determined from the upper critical field (see more detail on Sample $C$ in Appendix A).
Similarly, we use Eq.\ref{Eq:Idep0} together with the temperature dependence expressed in Eq.\ref{Eq:Idep}
as a best estimate for the theoretical depairing current. For further study, we selected the samples with
the highest ratio of $j_{c}/j_{dep}$ to analyse the photon response.

Biased near the critical current we observe voltage pulses quite similar to
those we routinely observe with the usual meander-type SSPD's. Fig.~\ref{fig:Figure2}(b)
presents a typical voltage transient of the 2.12-$\mu$m-wide bridge. The decay time is much shorter than in the meandering SSPDs, but it
is still longer than expected from the kinetic inductance $L_k$ of
the bridge \cite{KermanAPL2006} connected to the 3-$\Omega$ shunt
resistor. The value of $L_k$ in our samples is in the range of 0.4 to 1.1\,nH,
giving a characteristic decay time in the sub-nanosecond range. We
attribute this discrepancy to a parasitic inductance of the read-out
lines and the mounting.

\section{Single Photon Detection}

The photon-detection has been carried out in an experimental set-up discussed in detail in Appendix B. Fig.~\ref{fig:Pulses} presents
real-time waveform transients taken by a digital oscilloscope. The
top blue curve shows the clock pulses from the laser. All the other
red curves are the responses from the sample measured for decreasing power, by increasing
the optical attenuation. We observe: (1) the amplitude of the
photoresponse does not depend on the attenuation, and (2) the
probability to observe a response drops with the increase of
optical attenuation.

\begin{figure}[t]
    \center
    \includegraphics[width=0.5\textwidth] {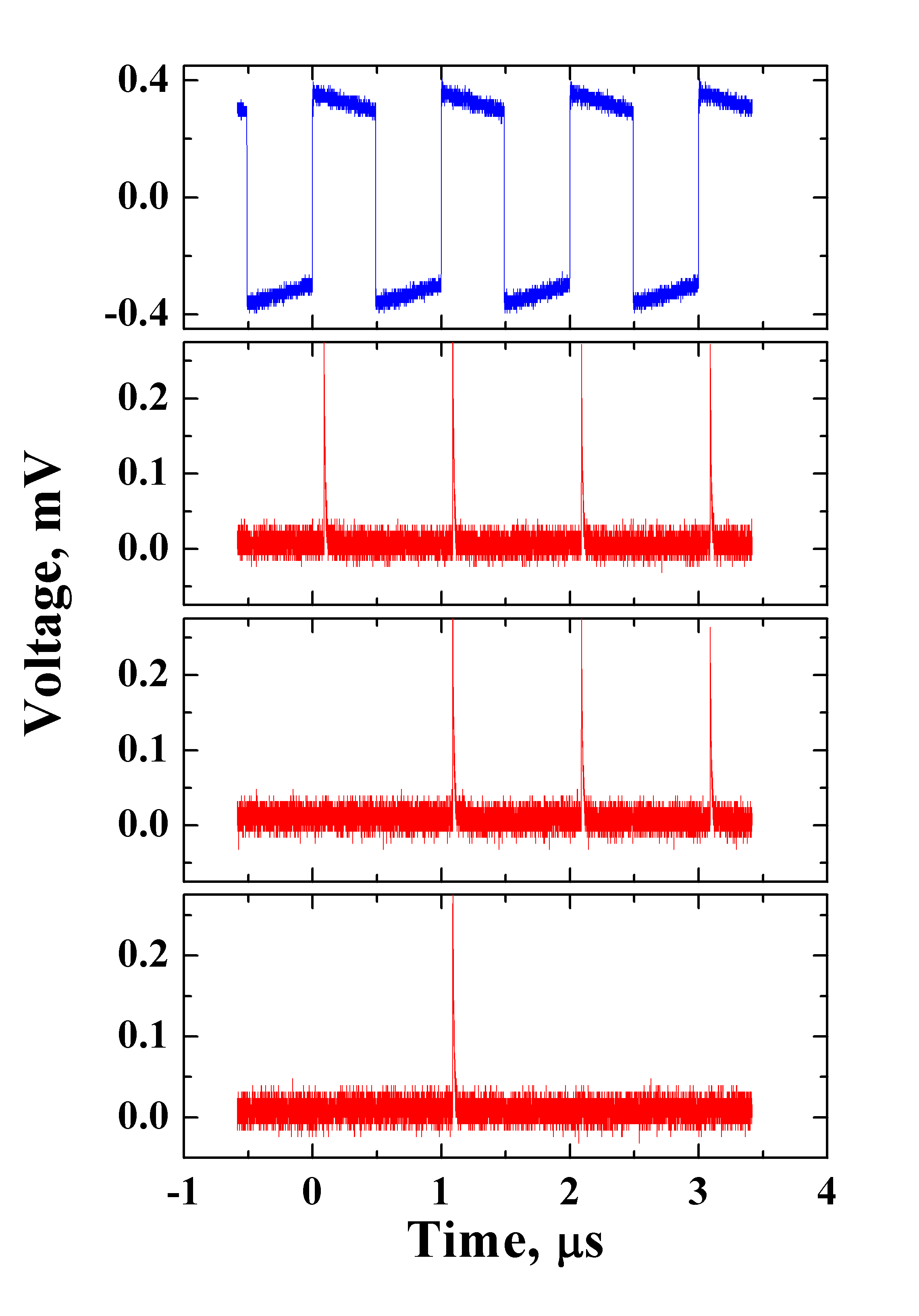}
    \caption{Real-time waveform record showing clock pulses from the laser (top blue)
    and photon pulses detected by the bridge at different attenuation levels of the power from the laser (red curves, power decreases from top to bottom). With the increase of attenuation of the power the number of detected pulses decreases. }
\label{fig:Pulses}
\end{figure}

\begin{figure}
        \includegraphics[width=0.5\textwidth] {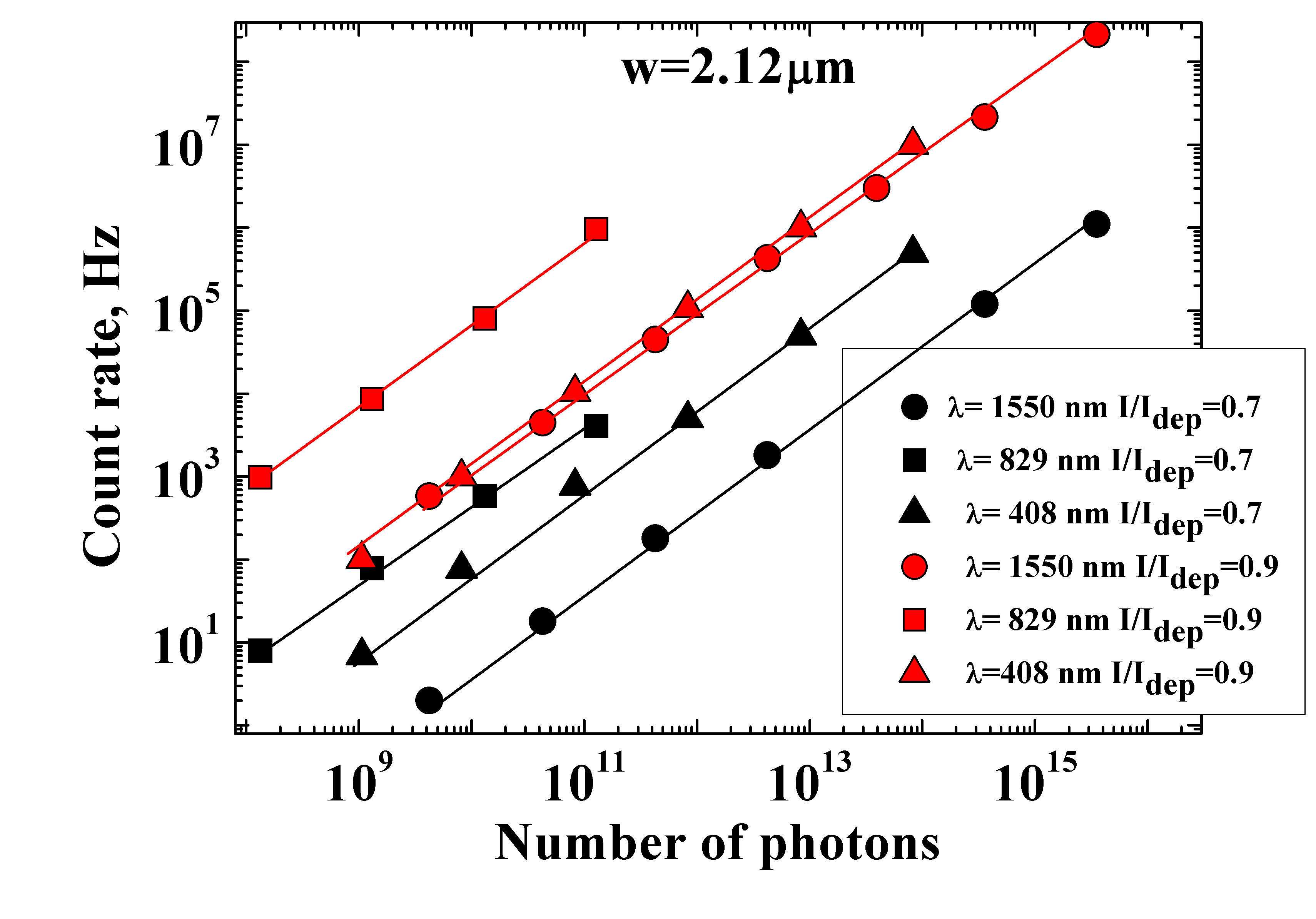} \\
\caption{Count rate \emph{versus} the number of photons in the laser pulse. Blue symbols for $I_{bias}=0.7I/I_{dep}$, red symbols for $I_{bias}=0.9I/I_{dep}$. The linear dependence of the count rate with the number of photons in the pulse corresponds to the Poisson statistics and indicates the single-photon nature of the response, irrespective of the bias current.
}
\label{fig:Poisson}
\end{figure}

\begin{figure}\center
\begin{tabular}{l}
    a)\\
    \includegraphics[width=0.5\textwidth]{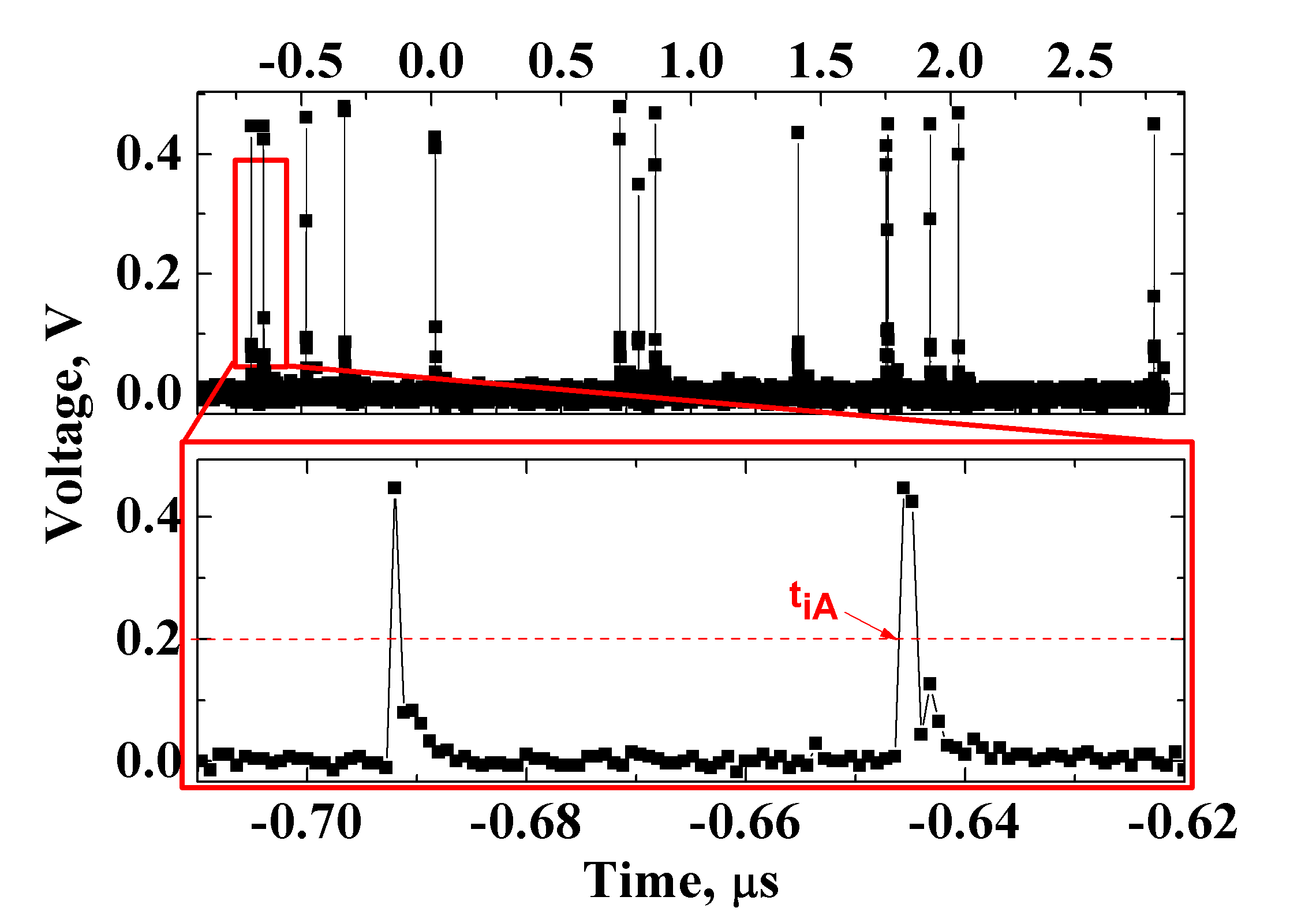} \\
    b) \\
    \includegraphics[width=0.5\textwidth]{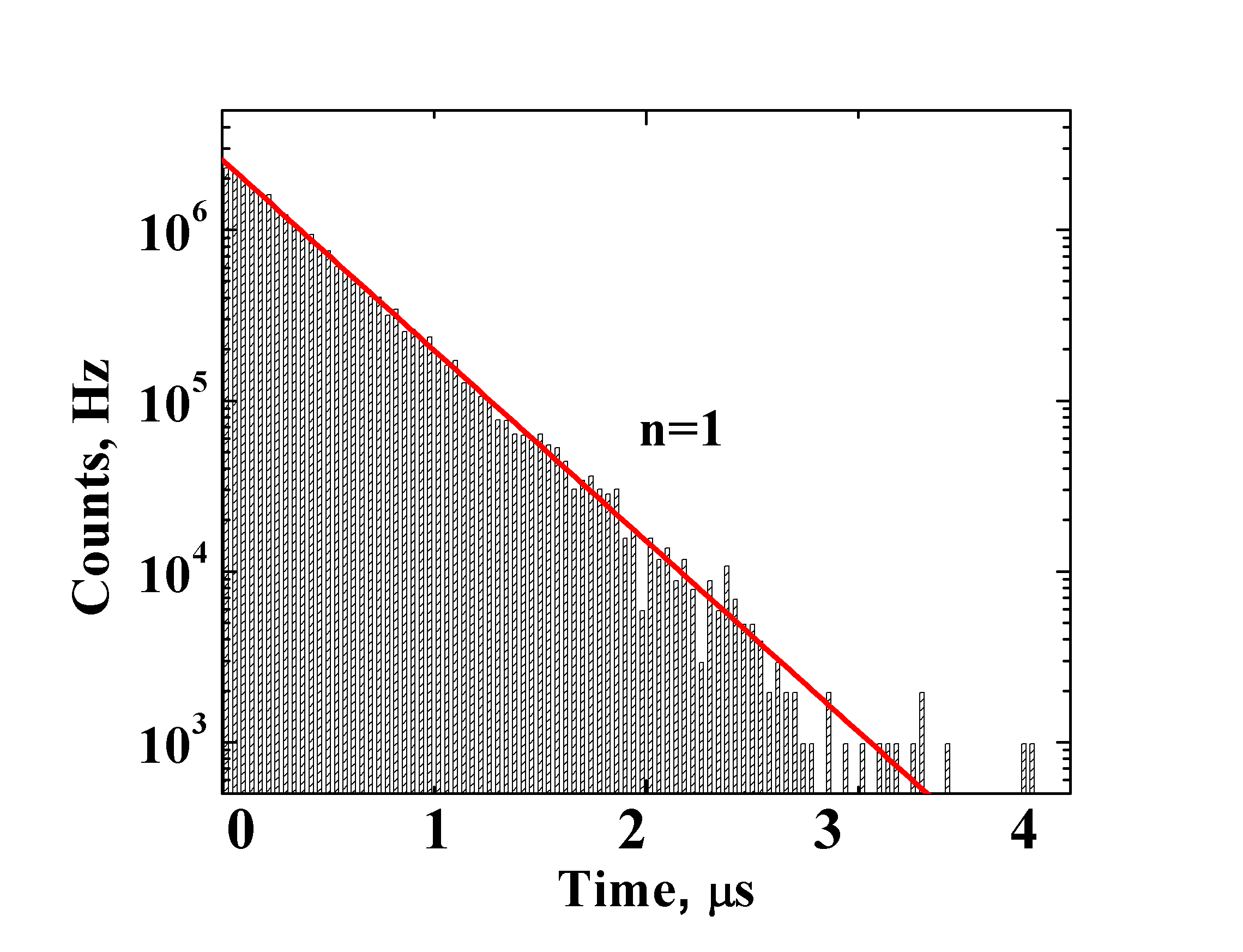}
\end{tabular}
\caption{(a) Oscilloscope waveform transient (top) and a fragment of the timetrace (bottom). (b) Statistical analysis of the inter-arrival time of the photon counts measured with cw laser of wavelength $\lambda$=1550 nm. The exponential distribution of the inter-arrival time intervals between photon counts shows the same Poisson distribution as for the photons in the incoming light providing additional proof of the single-photon response of the sample.
}
\label{fig:Figure23}
\end{figure}

We proceed with a statistical
analysis of the photon count rate on the number of
incident photons used previously in Refs. \citenum{GoltsmanAPL2001, KorneevaIEEE2017}.  In the single-photon counting regime the photon
count rate $R$ should be proportional to the photon flux
$R_{ph}$: $R\propto R_{ph}$. For multi-photon detection we expect that
$R\propto R_{ph}^n$, with $n$ the number of simultaneously
absorbed photons producing a single count. This behavior follows from the Poisson distribution of the incident photon-flux. In a strongly
attenuated laser beam the probability $p$ to have a given number
of photons $n$ in a given constant timeslot should be distributed
according to: $p\propto \langle m \rangle
\exp(-\langle m\rangle)/n!$, where $\langle m \rangle$ is the mean
number of photons in the timeslot. The probability $p$ of
detecting one photon is proportional to the mean photon number
$\langle m \rangle$, the probability of detecting two photons is
proportional $\langle m \rangle^2$, and so on. In
Fig.~\ref{fig:Poisson} we show the count rate \emph{vs} incident photon flux
for Sample $C$ at three wavelengths: 408\,nm, 829\,nm and 1550\,nm
and for 2 bias currents. We selected two bias currents which
are supposed to correspond to two different mechanisms of
photoresponse as proposed by Zotova and Vodolazov \cite{ZotovaSUST2014}: (1) \textit{Regime I} (at
$I_{bias}=0.5I/I_{dep}$) corresponds to the initial sharp increase
of the $IDE$ and (2) \textit{Regime II} (at $I_{bias}=0.78I/I_{dep}$)
denotes a much slower increase of $IDE$. More details about these
regimes will be given below. One can see that for all studied
wavelengths and bias currents we observe $\propto R_{ph}$, a
dependence which confirms the single-photon operation of the sample. We
observe the same results for all studied samples including
the largest  5-15$\mu$m-wide Sample $F$.

As a further test of the response we studied inter-arrival time
distribution of photon counts as suggested by Marsili et al
~\cite{MarsiliNanoLett2011}. The statistics of inter-arrival time
is studied using the digital oscilloscope Tektronix DPO--70404C.
We record the waveform-transient of maximum length, which is a total of 12.5
million points, covering
10 ms windows with 800 ps resolution. Such a time resolution made
it possible to obtain at least one point on the rising edge and
2-3 points on the decreasing edge of the pulse
(Fig.~\ref{fig:Figure23}(a)).
 As a result we have a set of times $t_i$ and, correspondingly, the instantaneous voltages $U_i$. Then, we extract all time-moments $t_{iA}$, which correspond to the appearance of photo-counts. As an objective criterion, we took the voltage rise above a threshold value, to indicate the voltage pulses $U_{iA}$, which count as events.  $U_{iA}$ is taken much larger than the noise amplitude. From the array $t_{iA}$ we determine the time intervals between all successive photo-counts: $\Delta t_i = t_{i+1}- t_i$. From this we construct a histogram of the distribution of these time intervals, normalized to the number of time intervals and their width. Fig.~\ref{fig:Figure23}(b) shows the histograms of
this inter-arrival time for Sample $D$.

\begin{figure*}
\begin{tabular}{ll}
a) & b) \\
\includegraphics[width=8cm]{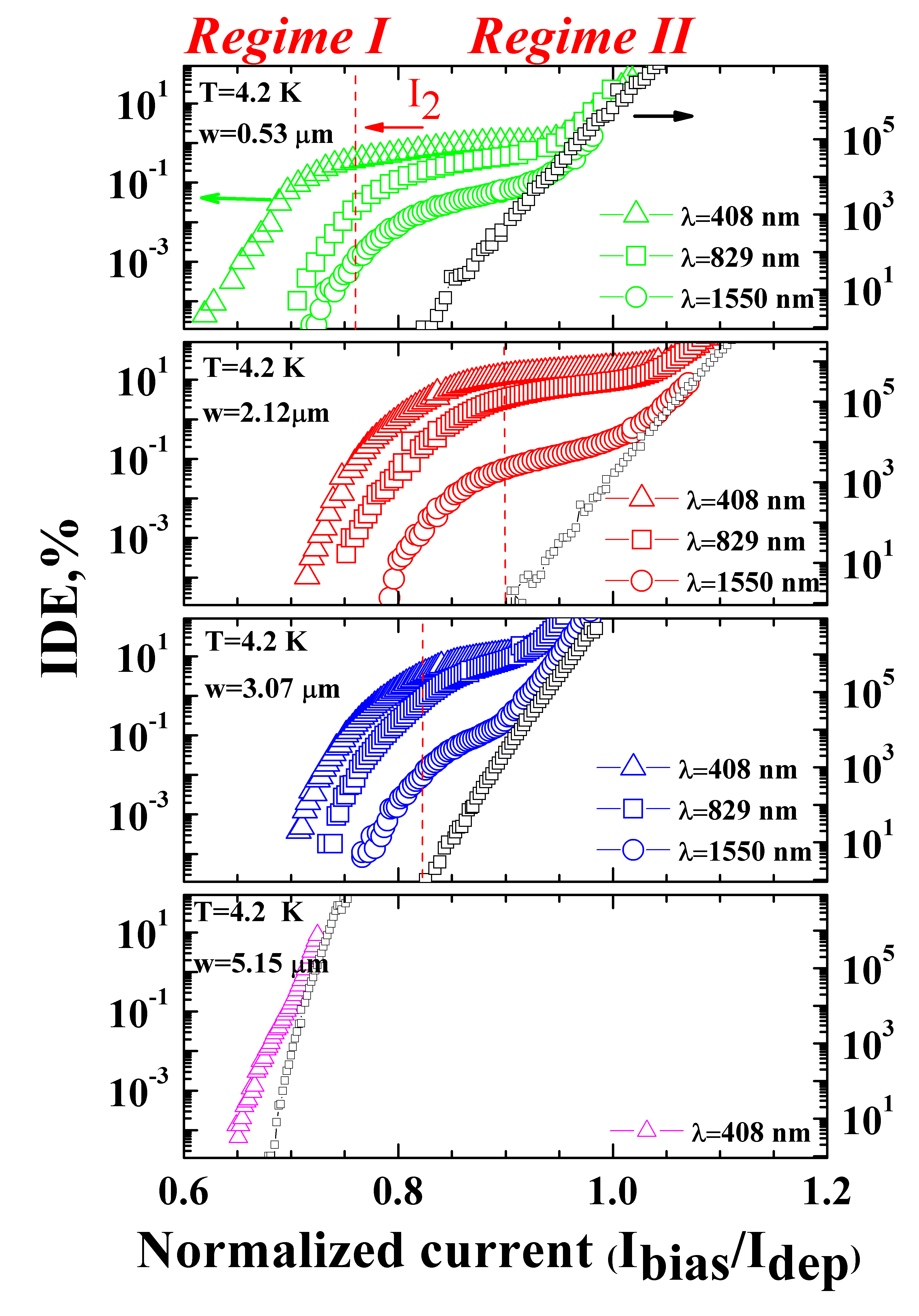} & \hspace{1cm} \includegraphics[width=8cm]{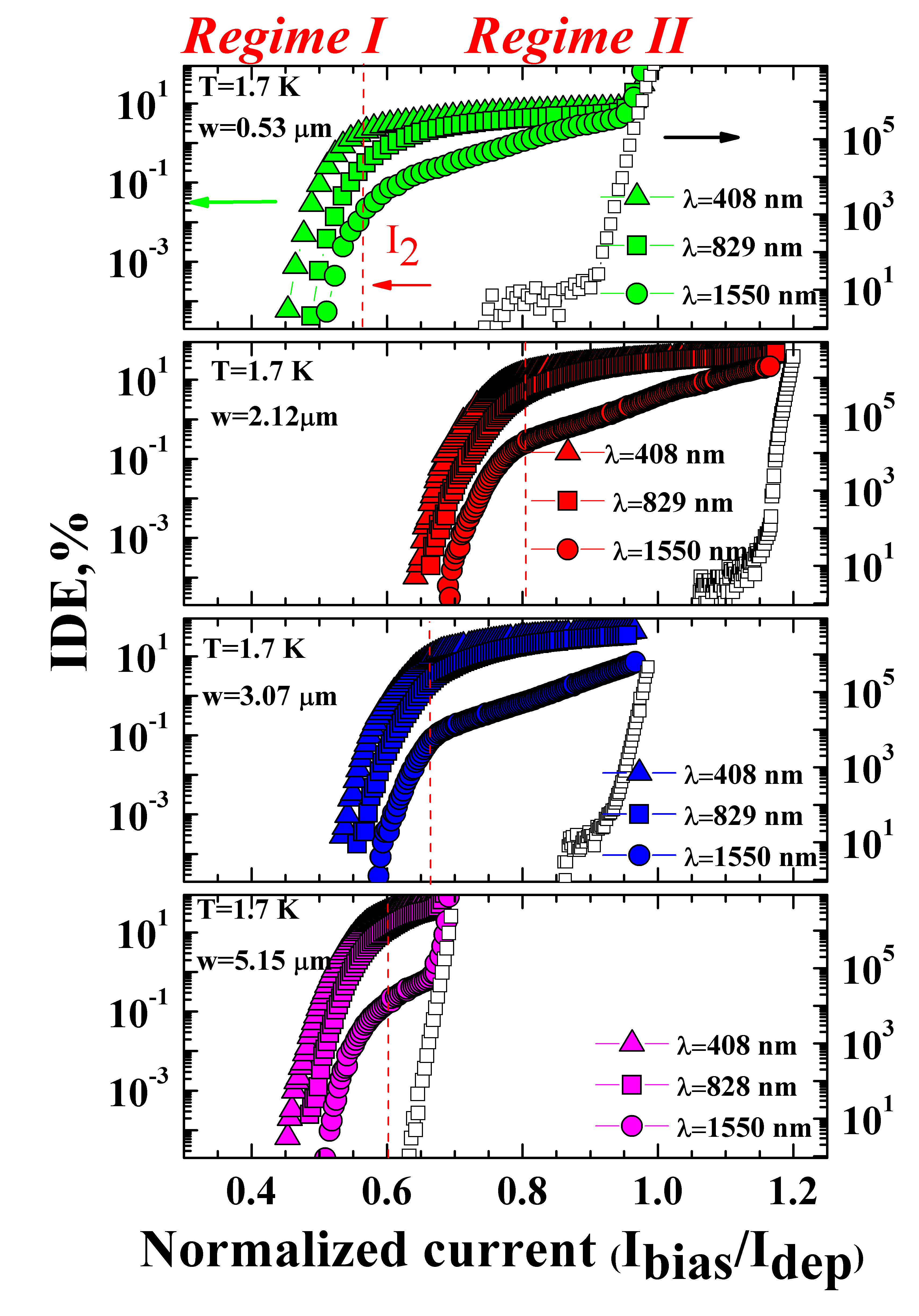}
\end{tabular}
\caption{Dependence of detection efficiency on bias current normalized to
the absorption, $IDE (I)$. The bias-current  $I_{bias}$ is normalized to the calculated depairing current $I_{dep}$.
(a) $IDE$ at 4.2\,K and (b) $IDE$ at 1.7\,K for the Samples (from top to bottom) $A$ with
$w$=0,53\,$\mu$m (green symbols), $C$ with $w$=2,12\,$\mu$m (red
symbols), $D$ with $w$=3,07\,$\mu$m (blue symbols) and $F$ with
$w$=5,15\,$\mu$m (magenta symbols). These dependencies show
two regimes: (1) with a sharp increase of the $IDE$ (\textit{Regime I})
and (2) with a much slower increase of the $IDE$ (\textit{Regime II}).}
\label{fig:IDE}
\end{figure*}

Since the photons in the incoming
light from the cw laser are independent and since they obey Poisson
statistics, the probability to record  $n$ photons in a time
interval $t$ is $ (\nu t)^n  \exp(-\nu t)/n!$, with $\nu$ the mean
photon flux. Let the first count being observed at  $t = 0$. The
probability of a second count during the interval from $t$ to
$t+dt$ is the multiplication of probabilities to have exactly one
photon in the interval $[t, t+dt]$ and $n-1$ photons in $[0, t]$.
The probability of the former event is $\nu dt$, the latter is
$(\nu t)^{n-1}  \exp(-\nu t)/(n-1)!$. Thus, the probability
distribution for the second photon count appearance is:
\begin{equation}
\rho(t)=\frac{\nu (\nu t)^{n-1}  \exp(-\nu t)}{(n-1)!}
\label{Eq:probability}
\end{equation}
The red straight line in Fig.\ref{fig:Figure23} is the prediction of
Eq.\ref{Eq:probability} with $n=1$ ($n$ is number of photons for a time
interval $t$). It clearly proves that the sample
does not accumulate more than one photon to produce a single photon count.

\section{Dependence of detection efficiency on current}

In Fig.~\ref{fig:IDE}(a,b) we show the evolution of the internal
detection efficiency $IDE$ with bias current. We distinguish two regimes, indicated in the figure. In \textit{Regime I} the $IDE$ grows fast in an exponential-like manner.
 This \textit{Regime I} we associate with fluctuation assisted photon counting. In this case, the absorption of a photon
triggers the transition to the resistive state only with the help
of thermally activated vortex nucleation near the point of impact
of the photon \cite{ZotovaSUST2014}. The indirect proof of this
intrinsic mechanism comes from the fact that the slopes of
$IDE$($I$) in \textit{Regime I} and the dependence of the dark
count rate on bias current (see Fig.~\ref{fig:IDE}(a,b)) are
identical for all samples and bias conditions.

The second regime, \textit{Regime II}, begins at a current denoted
by $I \gtrsim I_2$, which we relate to the position-dependent
photon counting proposed in
Refs.~\citenum{VodolazovPRA2017,ZotovaSUST2014,Engel2015}. In this
deterministic regime, the $IDE$ monotonically grows with the
current, starting from the current called $I_{det}^{min}$, where the region near the edges
of the sample (with typical width about the diameter
of the hot spot) starts to detect photons. It grows up to a current 
$I_{det}^{max}$, at which the central
part of the sample joins the detection process. For relatively
narrow widths it is expected that $I_{det}^{min}\simeq
I_{det}^{max}-0.1 I_{dep}$ (see Fig. 9 in
Ref~\citenum{VodolazovPRA2017}). The calculations for wider
samples give $I_{det}^{min}\simeq I_{det}^{max}-0.03\div 0.04
I_{dep}$ and dependence $IDE(I)$ has step like
form with $IDE \ll 1$ for $I<I_{det}^{min}$ and $IDE=1$ at
$I>I_{det}^{max}$ due to small contribution of near-edge region of
the sample to full intrinsic detection efficiency. Therefore we
can safely assume that current $I_2\sim I_{det}^{max}$. 

The model of Ref.~\cite{VodolazovPRA2017} predicts that the ratio
$I_{det}^{max}/I_{dep}$ increases at higher temperatures (see
Fig.\ref{fig:Calculation} in Appendix C) and so does
$I_2/I_{dep}$. This result qualitatively coincides with the
present experimental findings. At $T$=4.2\,K$\simeq 0.5T_c$ ratio
$I_2/I_{dep}$ has a larger value for all the bridges than at
$T$=1.7\,K $\simeq 0.2 T_c$, but the deterministic regime extends
over a wider current interval at lower temperatures. Moreover, for
example Sample F ($w=5.15 \mu m$) with the lowest reduced critical
current at $T$=4.2\,K, detects photons at high temperature only in
the fluctuation assisted regime, while at $T$=1.7\,K it operates
in the deterministic regime too.

In our experiment \textit{Regime II} extends over the  current
range $\sim 0.1-0.3 I_{dep}$ (depending on the temperature and the
specific sample), which is much larger than the theory
\cite{VodolazovPRA2017} predicts.
We also do not
observe a saturation of $IDE$. We believe that the main reason is
the geometry of our bridges, Fig.~\ref{fig:View}. The width of the
bridge increases when one moves from its center to the leads and
the local current density decreases. Therefore with increasing
current a larger (longer) part of the bridge participates in the
detection process and the $IDE$ grows monotonically until the
bridge switches to the resistive state at $I>I_{c}^{sh}$.

\section{Discussion}

Now, we summarize and explain our observations by using the concept of vortex-assisted
detection, introduced by Zotova and Vodolazov \cite{ZotovaPRB2012,
VodolazovPRA2017,ZotovaSUST2014}.  Because our samples have widths large compared to
the estimated hot spot diameter and to the coherence length, one
expects that the detection mechanism should be insensitive to the
width of the bridge, and only dependent on the supercurrent density $j$.

The model has one specific prediction.  For
sufficiently large width of the bridge, the onset of deterministic
detection is governed by the current density rather than by the
current. Physically, this means that the registration of a detection event, which starts
upon exceeding the critical current density (the critical supervelocity) \cite{ZotovaPRB2012,ZotovaSUST2014},
is sensitive only to the local density of the current (and to the
size and 'depth' of the spot, determined by the energy of absorbed
photon), but not to the distance between the hot-spot and the edge
of the strip. The latter requirement is only true if the strip is sufficiently wide, compared to the size of the hot spot.

What sufficiently wide means, is seen in
Fig.~\ref{fig:Calculation}  of Appendix C where the normalized
detection current $I_{det}^{max}/I_{dep}$, which is proportional
to the density $j_{det}$,  saturates near $w=100\xi$. In
our experiment, for most of the samples $w>100\xi$ holds, and we
identify $I_{det}^{max}$ with \textit{Regime II} with onset current
$I_2$. Hence, the density of this current $I_2/w=j_{det}$ for a given
wavelength is predicted to be the same for all the samples
(excluding maybe the narrowest sample A with a width of the order of
100 $\xi$ which is a borderline-case).

A direct check of this
prediction needs to take into account the following two aspects.  1) there is some on-chip resistance in series with the superconducting bridge, 
and 2) there is the shunt resistance
connected in parallel to the chip. Therefore,  we do not know the current,
flowing \emph{through the bridge}, with sufficient accuracy.
Hence, in the raw data in Fig.~\ref{fig:IDE}, one observes a spread of
\emph{measured} normalized detection currents $I_2/I_{dep}$ over
the different samples.

\begin{figure}\center
\begin{tabular}{l}
    a)\\
    \includegraphics[width=0.5\textwidth]{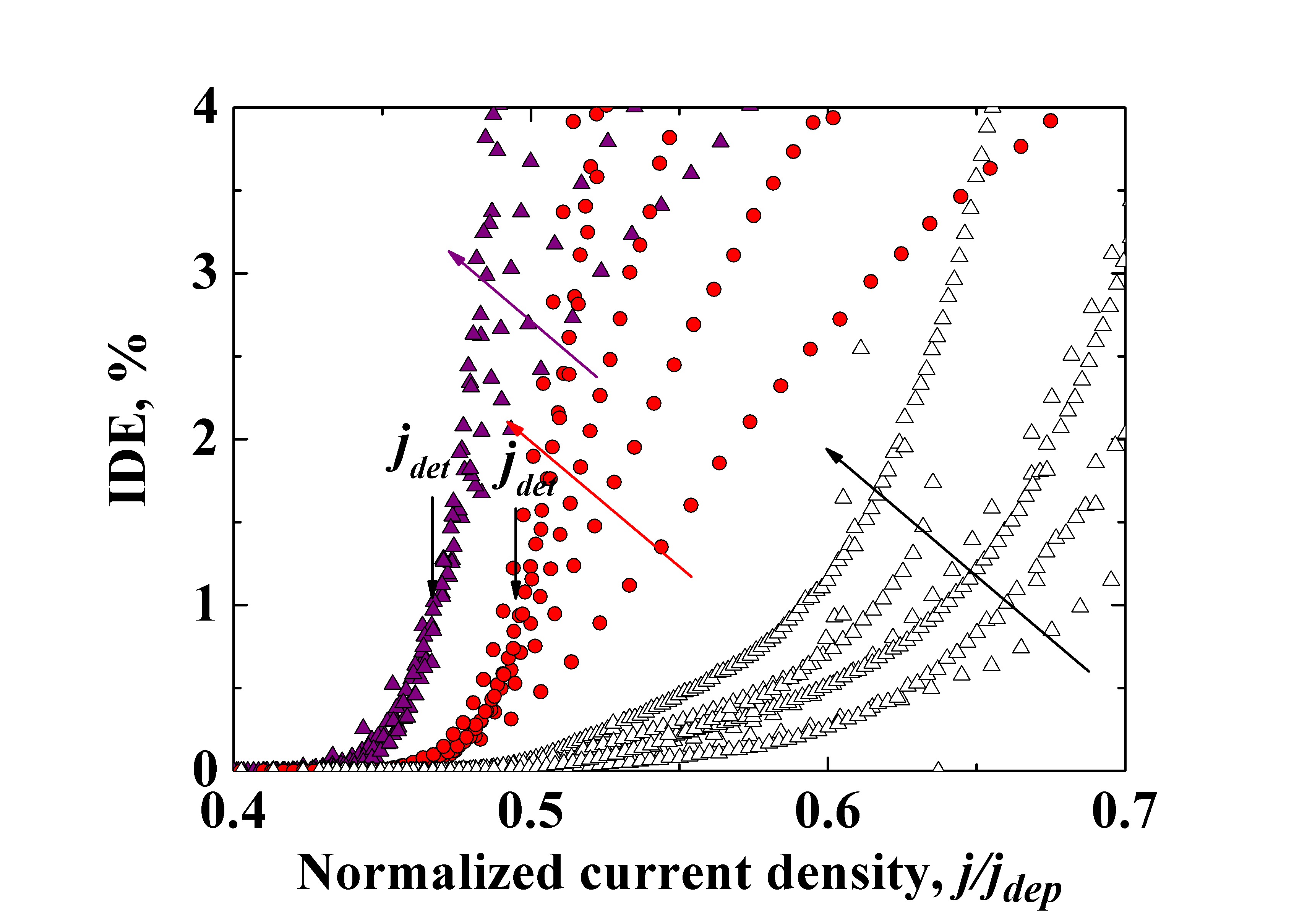} \\
    b) \\
    \includegraphics[width=0.5\textwidth]{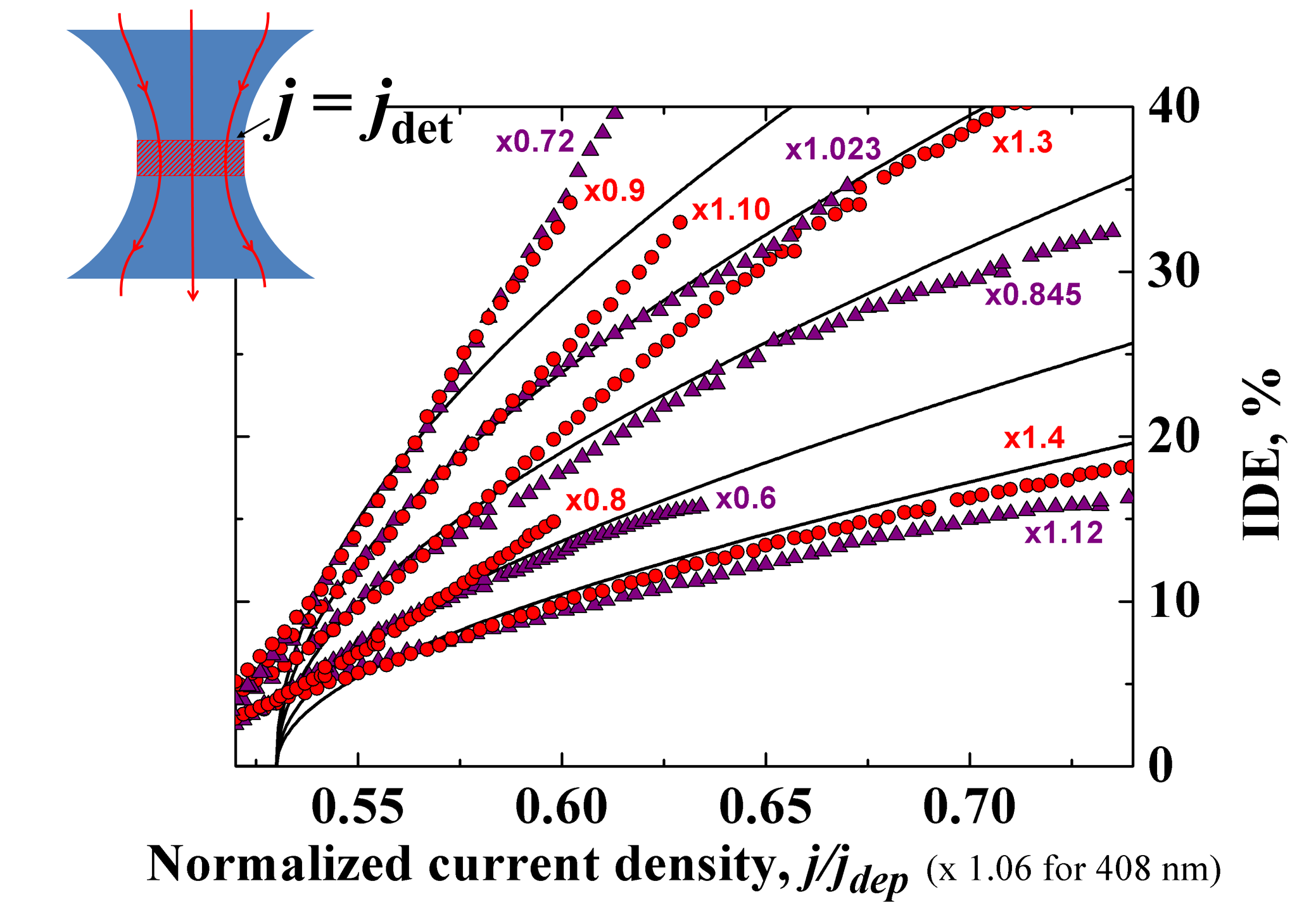}
\end{tabular}
\caption{\label{fig:Idet_invariance} (a) Dependencies of $IDE$ on current density near the on-set of \textit{Regime II}, demonstrating invariance of detection current density $j_{det}$ with respect to the width of the bridge. Violet, red and empty black symbols correspond to the wavelengths of 408\,nm, 829\,nm and 1550\,nm. Widths of the samples are $w$=0,53\,$\mu$m (Sample $A$), 1,06\,$\mu$m, 1,61\,$\mu$m, (Sample $B$) 2,12\,$\mu$m ($C$),  3,07\,$\mu$m ($D$) 4,07\,$\mu$m ($E$) and 5,15\,$\mu$m ($F$).  Arrows show how they correspond to the curves, from narrow to wide. The temperature is 1.7 K. (b) $IDE$ in the \textit{Regime II} for the wavelength of 408\,nm and 829\,nm, compared to ratio of detecting area of bridge to its total area. Data for samples $C$ - $F$ is shown, from bottom to top. The temperature is 1.7 K. Inset: schematics of bridge biased by a supercurrent. The part of the bridge, at which $j>j_{det}$, detects all absorbed photons, the outer part with $j<j_{det}$ does not.} 

\end{figure}

However, we find that the results for different samples are unaffected by the wavelength
of detected photons, despite of the fact we have sample to sample variations and an uncertainty in the the division of the current between the superconductor and the shunt. 
 This means that the ratio of the detection currents, 
\emph{even those measured with shunt}, corresponding
to two wavelength, $I_2(\lambda_1)/I_2(\lambda_2)$, should be the same for all samples with different widths. This prediction can be checked by 
renormalizing the bias current for all samples by dividing 
$I_2$ by the width $w$. This value should be the same for one wavelength. After
this re-normalization, we expect that $I_2/w$ at other wavelengths
will also be the same for all samples.

 Fig.~\ref{fig:Idet_invariance} a) demonstrates these results for a temperature of 1.7\,K.
 We normalized the bias currents to match $IDE(I)$ at $IDE=1$\,\% for the wavelength of 408\,nm. To relate
 the numbers to the current density, we apply the following procedure. We take the data for Sample $B$, 
 which has values closest to the critical current measured with and without shunt, hence presumably
 the lowest on-chip serial resistance. We then calculate the current flowing through the superconducting
 bridge by multiplying the measured current to the ratio $I_c/I_c^{sh}$ for this sample, and then divide
 by $I_{dep}$ at 1.7\,K to obtain the ratio $j/j_{dep}=I/I_{dep}$. One can clearly observe that the detection
 currents densities become very close to the wavelength of 829\,nm. This holds also for the wavelengths
 of 637\,nm and 937\,nm, not shown here. The only deviating sample is the
 narrowest Sample A, which is in line with the observations above.  One can also notice that for the largest
 wavelength of 1550\,nm the variation of the detection current density is significantly worse, which
 we relate to the breakdown of the pure deterministic detection at low energies of photons.

The same argument about the dependence of $IDE$
on $j$ rather than on $I$ holds also for fluctuation-assisted
detection. In this case, one assumes that the detection events in
this regime occur in the narrowest part of the bridge with the
highest current density. This is also consistent with the data.

Next, for wide strips of constant width, the
deterministic nature of detection should result in a step-like
dependence of $IDE$ in its dependence on $j$.  $IDE \ll 1$ at $j<j_{det}$ and $IDE=1$ at
$j>j_{det}$. For our neck-shaped samples,
we assume that we have unit probability of response to the
absorbed photon in the central part of bridge, where 
$j >j_{det}$.  The probability is zero further away from the narrowest part  (see inset on Fig. ~\ref{fig:Idet_invariance}). The boundary between the inner and outer parts is set by the
condition $j=j_{det}$. Hence, introducing the area of the central,
'detecting', part $S_{det}(j)$, we derive the prediction that the
internal detection efficiency is $IDE(j)=S_{det}(j)/S$, where $S$ is
the total area of the bridge. (The details on calculation of this
$IDE(j)$ can be found in the Appendix D). Fig. Fig.~\ref{fig:Idet_invariance} b) compares
this prediction with the experiment for the wavelengths of 408\,nm
and 829\,nm. One sees that the model reproduces the three
features: i) steeper increase of $IDE(j)$ for wider bridges, ii)
convex shape of the $IDE(j)$ dependencies, and iii) up to a factor
of 0.7 - 1.5, the absolute value of measured $IDE$. The last
discrepancy can be explained by systematic and stochastic errors
of our method of determination of $IDE$. The non-regular deviation
of the curves for some of the samples from the model prediction can
be attributed to defects in the samples. The model agrees with the experiment qualitatively,
and, even quantitatively for a sizable fraction of the samples.  This is a strong indication
that we observe photon detection with near-unity intrinsic
probability at the short wavelengths.

Comparing
these findings to the predictions of the microscopic theory, we note that the observed detection
current density ${j_det} \approx 0.5 {j_dep}$ for the wavelength
of 408\,nm is close to the calculated one
\citenum{VodolazovPRA2017}. As expected, $j_{det}$ increases with
the wavelength (\emph{i.e.} with the decrease of photon energy)
and appears to equal the experimental $j_c \approx 0.7 j_{dep}$ for
the wavelength of 1550\,nm. This means that, to realize IDE close to unity for near-IR photons, one has to reach either a larger $j_c/j_{dep}$, or, which seems more achievable, to enhance the effect of the hot spot, created by IR photon, on the current density, using thinner films.

\section{Conclusion}

We have developed single photon detectors based on NbN
microbridges. The dependence of the Internal Detection Efficiency
($IDE$) on the supercurrent qualitatively resembles those of
meander-type single photon detectors (SSPDs),  with widths less
than 200 nm. Our results deomostrate a new type of single photon
detector based on a short superconducting bridge with dimensions
comparable to the diameter of an optical fiber and
an IDE of about one. This design provides a much shorter dead
time, which is in the presently used detectors several
nanoseconds, due to the long length of the
meander-type nanowire. Indirectly our results confirm the vortex
assisted mechanism of photon detection by wide current-carrying
strip as originally proposed by one of the authors
\cite{VodolazovPRA2017}.


\appendix

\section{Device fabrication and characterization}

\begin{figure}
    \centering
    \begin{tabular}{l}
        a) \\
        \includegraphics[width=0.5\textwidth] {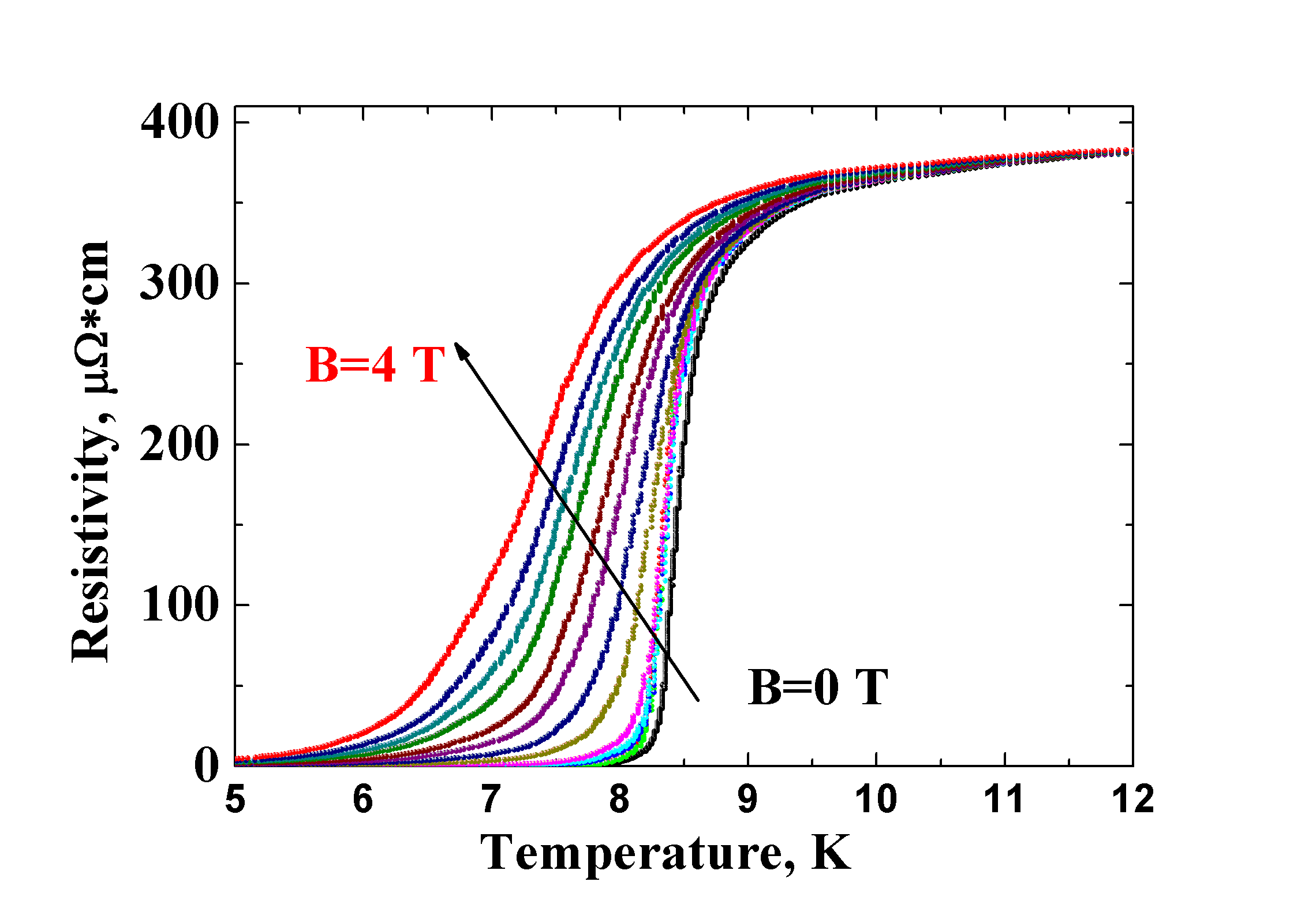}\\
        b) \\
        \includegraphics[width=0.5\textwidth]{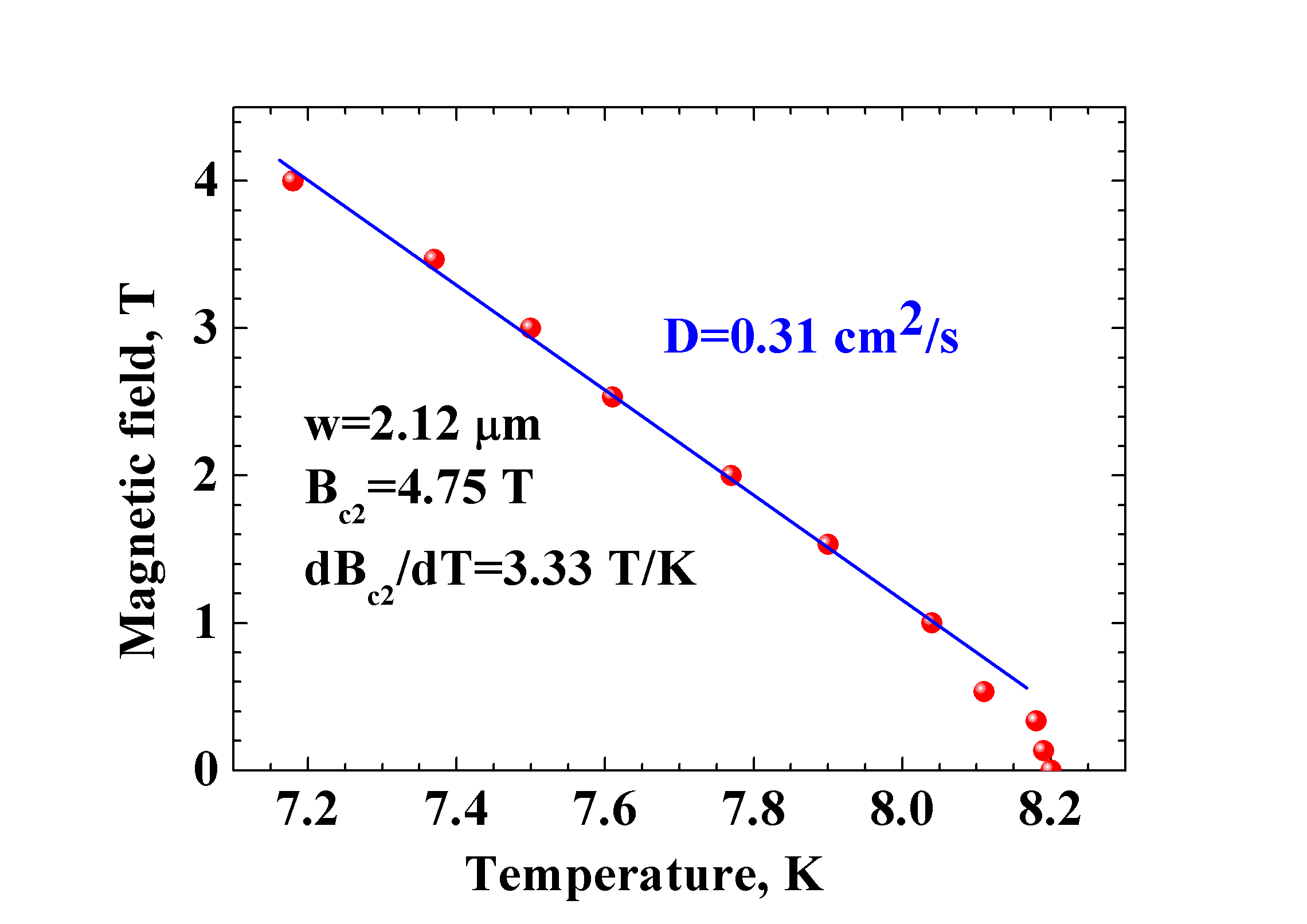}

    \end{tabular}
    \caption{(a) Dependence of resistivity on temperature for 2.12-$\mu$m-wide Sample $C$ at different magnetic fields in the range from $B$=0\,T to $B$=4\,T.
       The black arrow indicates the direction of increasing $B$. (b) Measured temperature dependence of the critical magnetic field for a 2.12-$\mu$m-wide Sample $C$ (circles), and linear fits of the data (solid line) used for $B_{c2}(0)$ and the determination of the diffusivity.}
    \label{fig:Diffusivity}
\end{figure}

The 5.8\,nm-thick  NbN film is deposited by dc  magnetron sputtering of a niobium target in a plasma
 consisting of a mixture of argon (Ar) and nitrogen ($N_{2}$). The film is deposited on a silicon wafer with
 a buffer layer of silicon dioxide. The SiO$_2$ layer is 250\,nm-thick. Before starting the sputtering
 process the substrate is heated to 400$^\circ$\,C. The film is characterized by critical temperature of
 approximately $T_{c}$=8.3\,K. The deposition is done in the gas mixture with flow rates 40\,cm$^{3}$/s
 and 6.6\,cm$^{3}$/s for Ar and $N_{2}$, respectively, and a current of 550\,mA. The deposition rate under
 these  conditions is 0.88\,{\AA}/s.

The NbN film is patterned into single bridge with rounded edges,
using the electron-beam lithography and reactive ion
etching technique. From one film we make a batch of samples
with different widths in the range from 0.53\,$\mu$m to
5.15\,$\mu$m. The size of each sample is determined with a SEM. All
bridges are characterized by the critical temperature determined from
superconducting transition. The I-V curves are determined at temperatures
of $T$=4.2\,K and $T$=1.7\,K. From a measurement  of the temperature dependence of the
second critical magnetic field $B_{c2}$, we infer a
diffusion constant $D$=0.31\,cm$^2$/s, Fig.~\ref{fig:Diffusivity}. This value is determined for one device
 and assumed to be identical for the other devices. Based on experiments on similar samples~\cite{Shcherbatenko2016}, we expect that this causes an error of at most 20\%. Extrapolating the linear temperature dependence of $B_{c2}$ near $T_{c}$ to $T=0$ we find $B_{c2}(0)$ =4.75\,T.

\begin{figure}\center
    \includegraphics[width=0.5\textwidth]{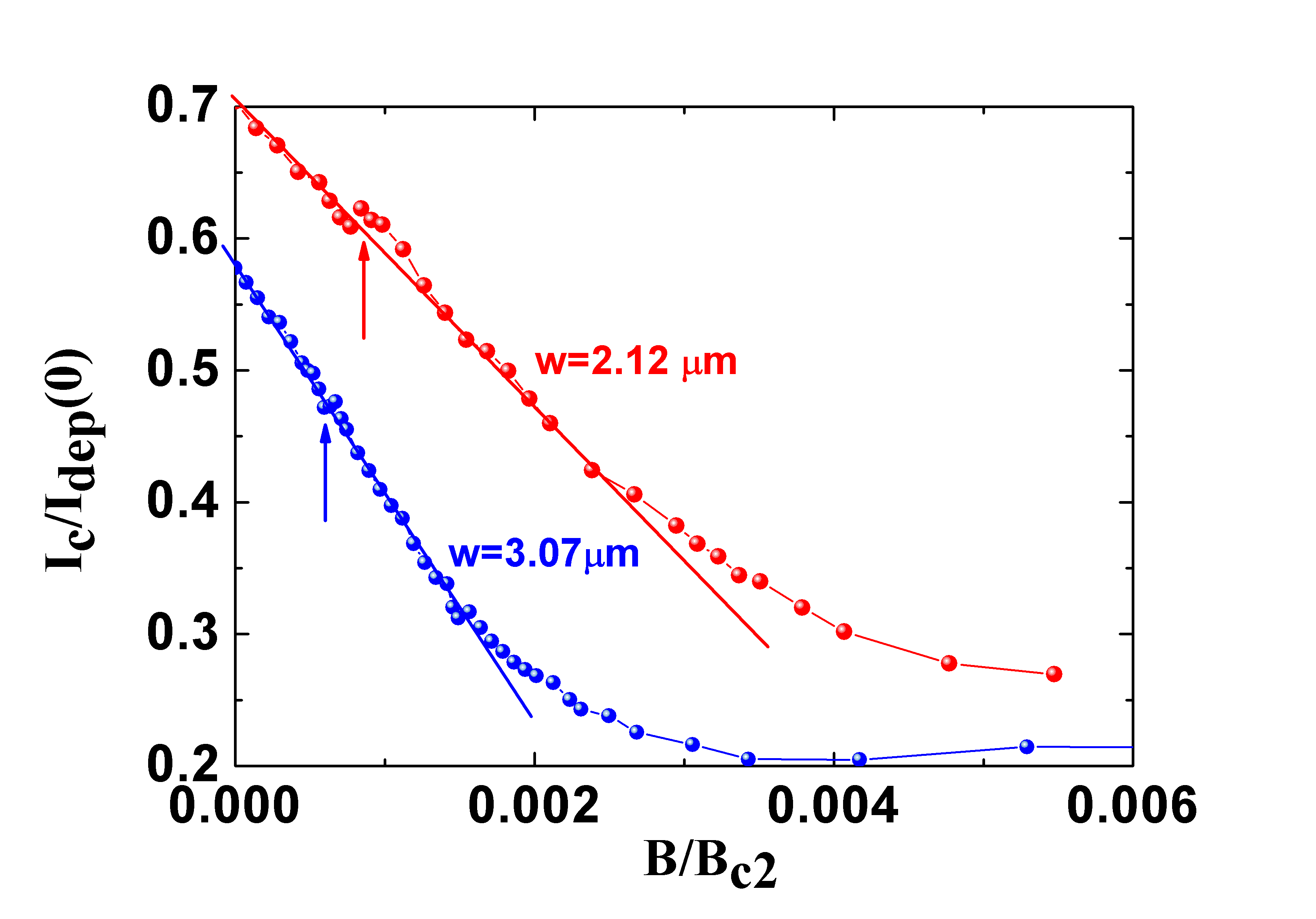}
    \caption{Dependence of the normalized critical current $I_c/I_{dep}(0)$ (without the shunt) on the magnetic field measured for 2-12 and 3-07$\mu$m-wide samples. Linear dependence in low magnetic fields is evidence that the NbN film does not have large extrinsic defects leading to vortex-pinning in the strip. The arrow-marked kinks are connected with the appearance of the single vortex chain in the middle of the bridge at that specific magnetic field strength. }
\label{fig:MagField}
\end{figure}

As an additional proof of high quality of our bridges we measure
dependence of critical current $I_c$ on perpendicular magnetic
field $B$. Fig.~\ref{fig:MagField} shows dependencies of $I_c$ on
$B$ measured for devices $C$ and $D$ (2.12- and 3.07-$\mu$m-wide).
We find the linear decay $I_c(B)$ at low magnetic field which
demonstrates dominant contribution of the edge barrier for vortex
entry \cite{PlourdePRB01} and kinks in dependence of $I_c(B)$ at
$B/B_{c2}\approx10^{-3}$ (marked by arrows) which are connected
with the presence of the single vortex chain in the middle of the
bridge \cite{ShmidtJETP1970,VodolazovPRB2013,IlinPhysRev2014}.
Both these results could not be observed in the bridge with
dominant contribution of bulk pinning to $I_c$ and presence of
large number of defects being able to pin vortices.

\section{Experimental set-up and measurements}

The electro-optical characterization of our samples is performed in a fiber-based set-up. The sample is mounted on a dipstick to be inserted into a liquid He dewar. The measurements are carried out at two temperatures: T=4.2\,K and T=1.7\,K. The latter is reached by vacuum pumping the helium from a cryo-insert for a storage dewar.
As light sources, we use light emitting diodes with wavelengths 408\,nm, 637\,nm, 829\,nm, 937\,nm, 1330\,nm and 1550\,nm which can be operated in both pulsed and cw regimes. The sample-chip with the transmission line is connected to a DC+RF-output port of a bias-T. The bias current is supplied through the DC port. We connect a 3\,$\Omega$ resistor in parallel to the sample to prevent latching when the critical current is exceeded. The voltage pulse is amplified by two room-temperature Mini-Circuits ZFL-1000LN+ (1-GHz band,46-dB total gain) amplifiers, and is fed to a digital oscilloscope and a pulse counter (Agilent 53131A (225 MHz band)).

\begin{figure}
    \centering
    \begin{tabular}{l}
            \includegraphics[width=0.5\textwidth] {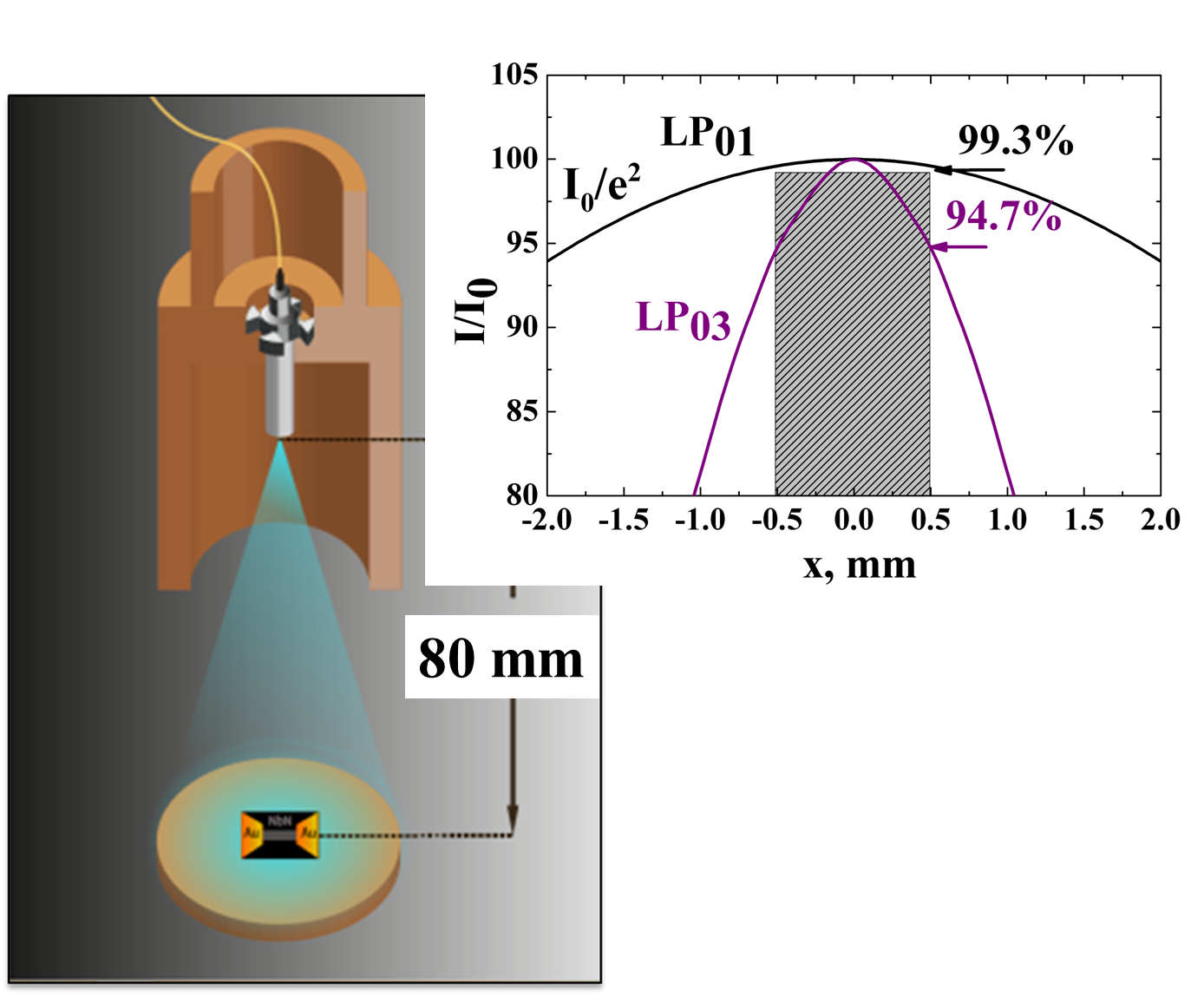}
          \end{tabular}
    \caption{A sketch of the sample illumination principle: to produce a uniform illumination of relatively large area we place the sample 80\,mm away from the fiber pigtail. The insert shows calculation of light beam intensity profile at the sample plane. Curve LP$_{01}$ is a single-mode profile which is observed in SM-28 fiber for wavelengths longer than 1\,$\mu$m. It is a Gaussian profile calculated as $I=I_{0}\exp(2x{^2}/r{^2})$, with $I_{0}$ is the irradiance at the center of the beam and $r_0$ is the radius of the beam at which the irradiance is $I_{0}/e^{2}$ and $r_{0}$= d$\cdot$NA, where $d$ is a distance from surface of optical fibre to surface of detector, $NA$ is numerical aperture. For fiber SM-28 $r_{0}$=11.2\,mm. Curve LP$_{03}$ corresponds to multi-mode profile at 408\,nm wavelength, it is described by Bessel functions of second kind. The gray rectangle corresponds to 0.5\,mm displacement, one can see that in the worst case the light intensity is at least 94\% of its maximum in the center.}
    \label{fig:SetUp}
\end{figure}

In view of the topology and the small active area of our
samples we do not package them with a single mode fiber as
usually done with meander SSPDs ~\cite{Miki2007, SlyszAPL2006}. For the present measurements we
use the sample holder, shown in Fig.~\ref{fig:SetUp}. In
this sample holder we use the optical fiber SM-28, which is single
mode for wavelength 1550\,nm with 9\,$\mu$m core diameter and a NA
(numerical aperture) of 0.14. However, at lower cut-off wavelength
(below 1260\,nm) the mode distribution in SM-28 fiber is not
Gaussian, because the fiber becomes multimode~\cite {Senior2009}.
To illuminate our bridge uniformly, we increased the diameter of
the output Gaussian beam by placing the sample at a distance
of d=80\,mm from the end of the fiber. Fig.~\ref{fig:SetUp}
shows the calculated field profiles emitted from the fiber for the wavelengths
1500\,nm and 408\,nm. At wavelengths of 1310\,nm and 1550\,nm the light intensity within
1-mm-diameter spot is not less than 99\% of the light intensity in the center (LP$_{01}$ profile
shown in Fig.~\ref{fig:SetUp}). For 408\,nm wavelength the mode
profile is narrower (LP$_{03}$ in Fig.~\ref{fig:SetUp}), but
even in this case if the device is displaced by $\pm$0.5\,mm
the light intensity is not less than 95\% of the intensity in the center.

The dipstick is calibrated with a meander-SSPD with a filling
factor of 50\% (100\,nm strip and 100\,nm gap). The $IDE$ of the meander
SSPD has been previously measured by packaging it with single mode fibers.
Subsequent measurements of this sample in the dipstick with known Detection
Efficiency makes it possible to determine the number of
photons in the flow incident on the sample with an area of
10$\mu$m$\times$10$\mu$m. Knowing the ratio of the areas of the
meander-shaped SSPD and the bridge we calculate the number of photons
incident on the bridge.

\section{Theoretical estimates}

\begin{figure}[t]
\includegraphics[width=0.5\textwidth]{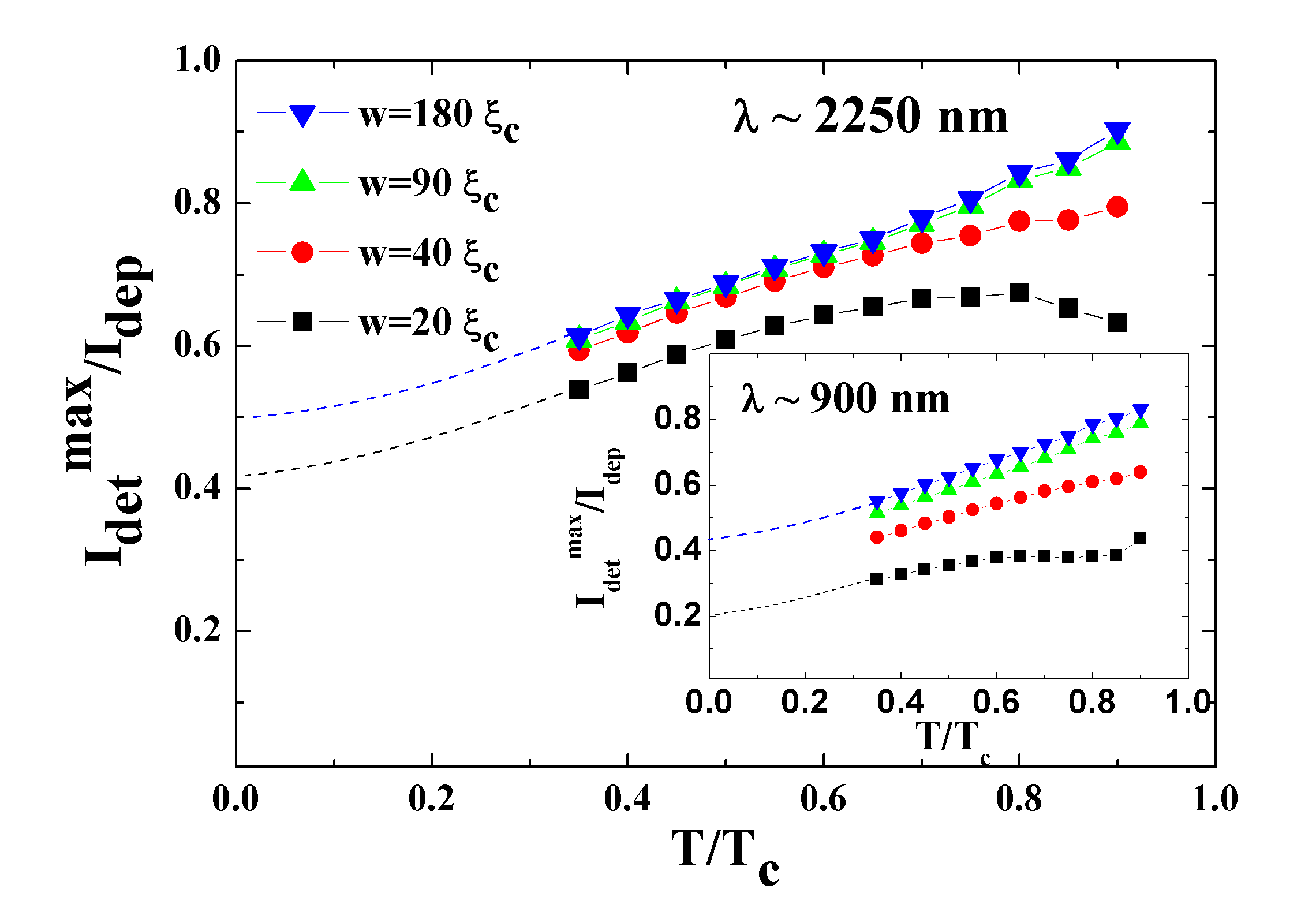}
\caption{\label{fig:Calculation} Dependence of
$I_{det}^{max}/I_{dep}$ on temperature for two wavelengths
$\lambda$=2250\,nm and $\lambda$=900\,nm (in insert) of the photon
and different widths of the strip. At the current $I \geq
I_{det}^{max}$ all points across the strip participate in photon
detection and intrinsic detection efficiency reaches unity.
Calculations are made in framework of 2T hot spot model from
Ref.~\citenum{VodolazovPRA2017}. Dashed lines show expected
dependence at low temperatures (they follow from results for hot
belt model -- see Fig. 6 in Ref.~\citenum{VodolazovPRA2017}) where
numerical scheme from Ref.~\citenum{VodolazovPRA2017} does not
converge.}
\end{figure}

In Fig.~\ref{fig:Calculation} we present the calculated dependence of
$I_{det}^{max}/I_{dep}$ on temperature for strips with different
widths and two wavelengths. The results are obtained in the framework of
the two-temperature hot spot model developed by one of the authors in Ref.~\citenum{VodolazovPRA2017}. The calculations have been carried out at
temperatures of $T\geq 0.35 T_c$,  where the numerical procedure convergences well. In this model $I_{det}^{max}$ is
defined as the maximal value of the detection current at which all
points across the strip participate in the photon detection (and
where the intrinsic detection efficiency reaches unity). The growth of
$I_{det}^{max}/I_{dep}$ with temperature in the range of range $0.35$ to $0.7 T_c$ is
connected with nonlinear temperature dependence of the electronic and
the phonon energies \cite{VodolazovPRA2017}. The growth of
$I_{det}^{max}/I_{dep}$ at temperature $T \gtrsim 0.7 T_c$ for
wide strips ($w>40 \xi_c$, $\xi_c=\sqrt{\hbar D/k_BT_c}\simeq 5.4
nm $ for our bridges) is explained by the rapid drop of $I_{dep}(T)$.
This leads to a reduced Joule heating in the superconductor when the first
vortex-antivortex pair nucleates inside the (non-equilibrium) hot spot and worsens
conditions for the appearance of a fully normal domain. This is due to the same
reason that $I_{det}^{max}/I_{dep}$ grows for a strip with $w=20 \xi_c$
and $\lambda= 620 nm$ at $T>0.85 T_c$. One would expect that with
the approach to $T_c$ photon detection becomes impossible
\cite{VodolazovPRA2017} because the normal domain cannot appear in the
strip.

In the calculations we assume that the escape time of the nonequilibrium
phonons to the substrate $\tau_{esc}$ is equal to the characteristic time
$\tau_0 \sim$ 270\,ps. Furthermore that the important parameter $\gamma=10$. Both are  typical values for NbN (for the definition of $\tau_0$ and
$\gamma$, see Ref.~\citenum{VodolazovPRA2017}). Choosing smaller
value of $\tau_{esc}$ (up to $0.1 \tau_0$) does hardly change
the obtained dependencies, at least for temperatures $T<0.8 T_c$
because the time for the nucleation of a normal domain $\delta t$
\cite{VodolazovPRA2017} does not exceed $0.1 \tau_0$. At $T=0.9
T_c$ the time $\delta t$ approaches $0.4 \tau_0$ for a strip with $w=180
\xi_c$ . A smaller $\tau_{esc}$ provides a large value of
$I_{det}^{max}/I_{dep}$.

\section{Account for non-rectangular shape of bridges}

\begin{figure}[t]
\includegraphics[width=0.5\textwidth]{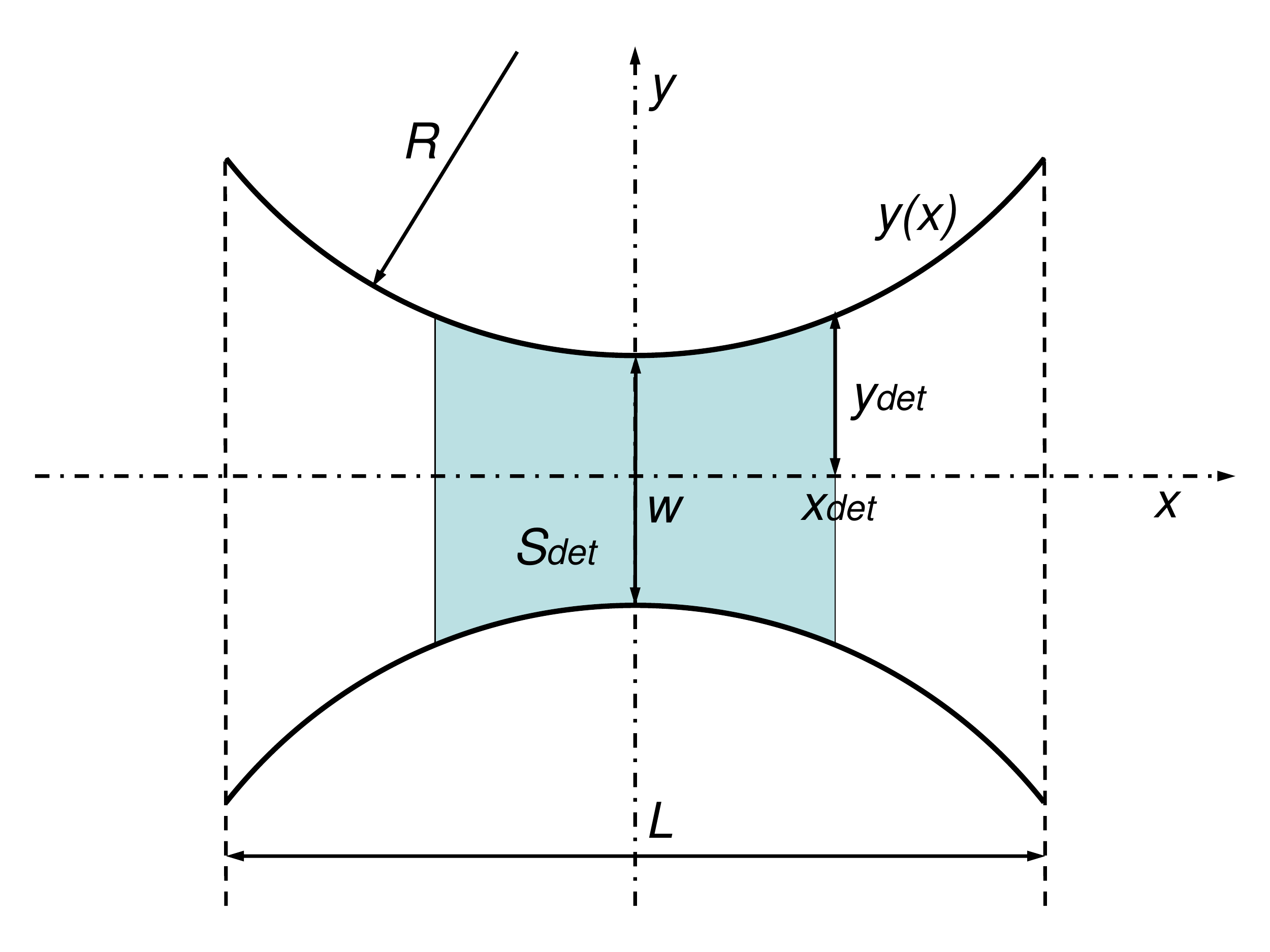}
\caption{\label{fig:S_det} On calculation of 'detecting' area $S_{det}$.}
\end{figure}

To apply our simple model of deterministic detection, which predicts $IDE=0$ at $j<j_{det}$ and $IDE=1$ at $j>j_{det}$, to our bridges with non-constant width and hence non-constant $j$ over the length, we calculate 'detecting' amount of the bridge area, in which $j>j_{det}$. To do it, we introduce coordinates as shown on Fig.~\ref{fig:S_det}. We express area of the bridge segment of length $2 x$ as $S(x)=4\int y(x) dx \approx 2wx+2x^3/3R$, where $y(x)=w/2+R-\sqrt{R^2-x^2} \approx w/2+x^2/2R$ is the half-width of the bridge at the cross-section with the coordinate $x$. The current density at the same cross-section is $j(x)=jw/y(x)$, with j the current density at the center of the bridge $x=0$. At the boundary between detecting and non detecting parts $x_{det}$, one has $j(x_{det})=j_{det}$ and $y(x_{det})=(j/j_{det})(w/2)$. Expressing half-width at this boundary as $y_{det} \approx Rw\sqrt{j/j_{det}-1}$ we derive, that the detecting area $S_{det}=S(x_{det}) \approx R^{1/2}w^{3/2}\left[2{(j/j_{det}-1)}^{1/2}+(2/3){(j/j_{det}-1)}^{3/2}\right]$. Finally, to obtain $IDE(j)=S_{det}/S$, we divide $S_{det}(j)$ to the total area of bridge $S \approx wL+L^3/12R$.

\begin{acknowledgements}

{\bf The work is supported} by the Russian Science Foundation (RSF) Project No.17-72-30036.  TMK is also supported by the European Research Council Advanced grant no. 339306 (METIQUM).
\end{acknowledgements}







\bibliography{allrefs2017}

\end{document}